\documentclass[10pt, twocolumn, final, twoside, journal]{IEEEtran}

\newcommand{\Descr}{\mathcal{D}}
\newcommand{\Descrt}{\tilde{\Descr}}

\newcommand{\descr}{d}
\newcommand{\descrq}{\tilde{\descr}}

\newcommand{\dv}{\mathbf{\descr}}
\newcommand{\dq}{\tilde{\dv}}

\newcommand{\cv}{\mathbf{c}}
\newcommand{\cq}{\tilde{\cv}}

\newcommand{\Dim}{P}

\newcommand{\scale}{\sigma}

\newcommand{\orient}{\theta}
\newcommand{\coord}{\mathbf{p}}
\newcommand{\coordq}{\tilde{\coord}}

\newcommand{\Rate}{R}

\newcommand{\Image}{\mathcal{I}}

\newcommand{\bag}{\mathbf{v}}
\newcommand{\recbag}{\mathbf{\tilde{v}}}

\newcommand{\qbagdex}{\tilde{q}}

\newcommand{\T}{\mathbf{T}}
\newcommand{\ATC}{\texttt{ATC}}
\newcommand{\ATCfull}{\textit{``Analyze-Then-Compress''}}
\newcommand{\CTA}{\texttt{CTA}}
\newcommand{\CTAfull}{\textit{``Compress-Then-Analyze''}}

\newcommand{\Dict}{\mathcal{V}}
\newcommand{\numvw}{V}
\newcommand{\vw}{\mathbf{w}}

\usepackage[shortlabels]{enumitem}

\usepackage{color, soul}
\definecolor{lightblue}{rgb}{.40,.90,1}
\sethlcolor{lightblue}

\usepackage{color}

\usepackage[pdftex]{graphicx}

\usepackage[table]{xcolor}
\usepackage[cmex10]{amsmath}
\usepackage{amssymb}
\usepackage{nicefrac}
\usepackage[tight,footnotesize]{subfigure}
\usepackage{multirow}

\usepackage{mathtools}

\begin{document}
\title{Coding local and global binary visual features extracted from video sequences}

\author{Luca Baroffio, Antonio Canclini, Matteo Cesana, Alessandro Redondi, Marco Tagliasacchi, Stefano Tubaro 
\thanks{Dipartimento di Elettronica, Informazione e Bioingegneria, Politecnico di Milano, Milan, Italy, email: name.surname@polimi.it}
\thanks{The project GreenEyes acknowledges the financial support of the Future and Emerging Technologies (FET) programme  within the Seventh Framework Programme for Research of the European Commission, under FET-Open grant number: 296676.}
\thanks{The material in this paper has been partially presented in \cite{Baroffio_VideoBRISK_ICIP14}.}
\thanks{Copyright (c) 2013 IEEE. Personal use of this material is permitted. However, permission to use this material for any other purposes must be obtained from the IEEE by sending a request to pubs-permissions@ieee.org.}
}

\maketitle

\begin{abstract}

	% Global and local image features are powerful tools that are exploited in several applications such as large scale visual search, object recognition and tracking, etc. 
	Binary local features represent an effective alternative to real-valued descriptors, leading to comparable results for many visual analysis tasks, while being characterized by significantly lower computational complexity and memory requirements. When dealing with large collections, a more compact representation based on global features is often preferred, which can be obtained from local features by means of, e.g., the Bag-of-Visual Word (BoVW) model. 
	Several applications, including for example visual sensor networks and mobile augmented reality, require visual features to be transmitted over a bandwidth-limited network, thus calling for coding techniques that aim at reducing the required bit budget, while attaining a target level of efficiency. In this paper we investigate a coding scheme tailored to both local and global binary features, which aims at exploiting both spatial and temporal redundancy by means of intra- and inter-frame coding. In this respect, the proposed coding scheme can be conveniently adopted to support the $\ATCfull$ ($\ATC$) paradigm. That is, visual features are extracted from the acquired content, encoded at remote nodes, and finally transmitted to a central controller that performs visual analysis. This is in contrast with the traditional approach, in which visual content is acquired at a node, compressed and then sent to a central unit for further processing, according to the $\CTAfull$ ($\CTA$) paradigm. 
In this paper we experimentally compare $\ATC$ and $\CTA$ by means of rate-efficiency curves in the context of two different visual analysis tasks: homography estimation and content-based retrieval. Our results show that the novel $\ATC$ paradigm based on the proposed coding primitives can be competitive with $\CTA$, especially in bandwidth limited scenarios.
\end{abstract}


\section{Introduction}
\label{sec:intro}
Visual analysis is often performed extracting a feature-based representation from the raw pixel domain. Indeed, visual features are being successfully exploited in a broad range of visual analysis tasks, ranging from image/video retrieval and classification, to object tracking and image registration. They provide a succinct, yet effective, representation of the visual content, while being invariant to many transformations. 

Several visual analysis applications (e.g., distributed monitoring and surveillance in visual sensor networks, mobile visual search and augmented reality, etc.) require visual content to be transmitted over a bandwidth-limited network. The traditional approach, denoted hereinafter as $\CTAfull$ ($\CTA$), consists in the following steps: the visual content is acquired by a sensor node in the form of still images or video sequences; then, it is encoded and efficiently transmitted to a central unit where visual feature extraction and analysis takes place. The central unit relies on a lossy representation of the acquired content, potentially leading to impaired performance. Furthermore, such a paradigm might lead to an inefficient management of bandwidth and storage resources, since a complete pixel-level representation might be unnecessary.  

In this respect, $\ATCfull$ ($\ATC$) represents an alternative approach to visual analysis in a networked scenario. Such a paradigm aims at moving part of the analysis from the central unit directly to sensing nodes. In particular, nodes process visual content in order to extract relevant information in the form of visual features. Then, such information is compressed and sent to a central unit, where visual analysis takes place.
The key tenet is that the rate necessary to encode visual features in $\ATC$ might be less than the rate needed for the original visual content in $\CTA$, when targeting the same level of efficiency in the visual analysis. This is particularly relevant in those applications in which visual analysis requires access to video sequences. Therefore, in order to maximize the rate saving, it is necessary to carefully select suitable visual features and design efficient coding schemes. 

In this paper we consider the problem of encoding both local and global binary features extracted from video sequences. The choice of this class of visual features is well motivated from different standpoints~\cite{CancliniCRTAC:2013}. First, binary features are significantly faster to compute than real-valued features such as SIFT~\cite{DBLP:journals/ijcv/Lowe04} or SURF~\cite{DBLP:conf/eccv/BayTG06}, thus being suitable whenever energy resources are an issue, such as in the case of low-power devices, where they constitute the only available option.
%and the only available option in low-power devices. 
Second, binary features have been recently shown to deliver performance close to state-of-the-art real-valued features. Third, 
%binary feature 
they can be compactly represented and coded with just a few bits~\cite{RedondiBACT:ICIP2013}. Forth, binary features are faster to match, thus being suitable when dealing with large scale collections. 

The processing pipeline for the extraction of local features comprises: i) a keypoint detector, which is responsible for the identification of a set of salient keypoints within an image, and ii) a keypoint descriptor, which assigns a description vector to each identified keypoint, based on the local image content. Within the class of local binary descriptors, BRIEF~\cite{DBLP:conf/eccv/CalonderLSF10} computes the descriptor elements as the result of pairwise comparisons between (smoothed) pixel intensity values that are randomly sampled from the neighborhood of a keypoint. BRISK~\cite{DBLP:conf/iccv/LeuteneggerCS11}, FREAK~\cite{DBLP:conf/cvpr/AlahiOV12} and ORB~\cite{Rublee_ORB} are inspired by BRIEF, and similarly to their predecessor, are also based on pairwise pixel intensity comparisons. They differ from each other in the way pixel pairs are spatially sampled in the image patch surrounding a given keypoint
%interest point. 
In particular, they introduce ad-hoc spatial patterns that define the location of the pixels to be compared. Furthermore, differently from BRIEF, they are designed so that the generated binary descriptors are scale- and rotation- invariant. More recently, in order to bridge the gap between binary and real-valued descriptors, BAMBOO~\cite{BAMBOO_MMSP}\cite{BAMBOO_ICASSP} adopts a richer dictionary of pixel intensity comparisons, and selects the most discriminative ones by means of a boosting algorithm. This leads to a matching accuracy similar to SIFT, while being 50x faster to compute. A similar idea is also exploited by BinBoost~\cite{Trzcinski13a}, which proposes a boosted binary descriptor based on a set of local gradients. BinBoost is shown to deliver state-of-the-art matching accuracy, at the cost of a computational complexity comparable to that of real-valued descriptors such as SIFT or SURF. 

On the other hand, global features represent a suitable alternative to local features when considering scenarios  in which very large amounts of data have to be processed, stored and matched. Global features computed by summarizing local features into a fixed-dimensional feature vector have been effectively employed in the context of large scale image and video retrieval~\cite{Jegou:2010:IBL:1718320.1718326}. Global features can be computed based on the Bag-of-Visual-Words (BoVW)~\cite{DBLP:conf/iccv/SivicZ03} model, which is inspired by traditional text-based retrieval.  VLAD~\cite{JDSP10} and Fisher Vectors~\cite{Perronin07_Fisher} represent more sophisticated approaches that achieve improved compactness and matching performance. More recently, the problem of building global features starting from sets of binary features was addressed in~\cite{Galvez11_BoBW} and~\cite{ICIP2014:Steinbach:BVLAD}, extending, respectively, the BoVW and VLAD model to the case of local binary features.  %RIFERIMENTO ICIP 2014 - ok
Solutions based on global image descriptors offer a good compromise between efficiency and accuracy, especially considering large scale image retrieval and classification. Nonetheless, local features still play a fundamental role, being usually employed to refine the results of such tasks~\cite{DBLP:conf/cvpr/PhilbinCISZ07}~\cite{DBLP:conf/iccv/SivicZ03}. Furthermore, the approaches based on global features disregard the spatial configuration of the keypoints, preventing the use of spatial verification mechanism and thus being unsuitable to tracking and structure-from-motion scenarios~\cite{DBLP:conf/eccv/JegouDS08, Fischler:1981:RSC:358669.358692}.

This paper proposes a number of novel contributions: 
\begin{enumerate}
	\item We consider the problem of coding local binary features extracted from video sequences, by exploiting both intra- and inter-frame coding. In this respect, we adopt the general architecture of our previous work~\cite{BaroffioRCTT:TIP}, which targeted real-valued features, and propose coding tools specifically devised for binary features.
	
	\item For the first time, we consider the problem of coding global binary features extracted from video sequences, obtained by summarizing local features according to the BoVW model, exploiting both intra- and inter-frame coding.
	
	\item We evaluate the proposed coding scheme in terms of rate-efficiency curves for two different visual analysis tasks: homography estimation and content-based retrieval. We show the impact of the main configuration parameters, namely, the number of keypoints, descriptor elements and visual words. Unlike our previous work, content-based retrieval is evaluated by means of a complete image retrieval pipeline, in which a video is used to query an image database. 
	
	\item We compare the overall performance of $\ATC$ vs. $\CTA$ for both analysis tasks. In the case of homography estimation, we show that $\ATC$ based on local features always outperforms $\CTA$ by a large margin. In the case of content-based retrieval, we show the $\ATC$ achieves a significantly lower bitrate than $\CTA$ when using global features, while it is on a par with $\CTA$ when using local features. 
	
\end{enumerate}
% 
% i) 
% 
% 
% a coding architecture suitable for binary local features extracted from video content and ii) an intra-/inter-frame coding architecture tailored to BoVW obtained aggregating binary descriptors. Both coding efficiency and accuracy performance on different visual analysis tasks are thoroughly evaluated by means of rate-accuracy curves, demonstrating the competitiveness of the ATC paradigm, based on binary local features, with respect to the traditional CTA paradigm, based on H.264/AVC video coding and the extraction of SIFT local features at the sink node. This work extends our previous work~\cite{Baroffio_VideoBRISK_ICIP14}, in which we illustrate the coding architecture tailored to local binary descriptors. In this paper we introduced a novel approach for computing and coding global image descriptors based on binary local features. Furthermore, we propose a new test scenario, namely Content Based Retrieval, in order to evaluate the effectiveness of both approaches.

% Two ad-hoc MPEG groups, namely CDVS and CDViS~\cite{CDVS}, are working towards the standardization of coding architectures tailored to local features extracted from still images and video sequences, respectively. 

In the context of local visual features, several past works tackled the problem of compressing both real-valued and binary local features extracted from still images. As for real-valued local features, architectures based on closed-loop predictive coding~\cite{icassp11:featuresorting}, transform coding~\cite{Chandrasekhar_transformcoding}\cite{RedondiCT:MMSP2012} and hashing~\cite{Torralba2008} were proposed. In this context, an ad-hoc MPEG group on Compact Descriptors for Visual Search (CDVS) has been working towards the definition of a standard~\cite{CDVS} that relies on SIFT features.
As for binary local features, predictive coding architectures aimed at exploiting either inter-descriptor correlation~\cite{Ascenso2013} or intra-descriptor redundancy~\cite{RedondiCT:ICIP2012} were proposed. Furthermore, Monteiro et al. proposed a clustering based coding architecture tailored to the context of binary descriptors~\cite{MonteiroPCS2013}.
%
% Furthermore, 
Moreover, some works aimed at modifying traditional extraction algorithms, so that the output data is more compact or more suitable for compression. In this context, CHOG~\cite{DBLP:conf/cvpr/ChandrasekharTCTGG09} is a gradient-based descriptor that offers performance comparable to that of SIFT at a much lower bitrate.
As an alternative approach, Chao et al.~\cite{ICIP2011:Chao::PreservingSIFT} studied how to adjust the JPEG quantization matrix in order to preserve local features extracted from decoded images. 

% modificare paragrafo

The problem of encoding visual features extracted from video content has been addressed only very recently. Makar et al.~\cite{Makar2012, Makar2013} propose to encode and transmit temporally coherent image patches in the pixel-domain, for augmented reality and image recognition applications. Thus, the detector is applied at the transmitter side, while the descriptors are extracted from decoded patches at the receiver. The encoding of local features (both keypoint locations and descriptors) extracted from video sequences was addressed for the first time in~\cite{BaroffioRCTT:ICIP2013} for the case of real-valued features (SIFT and SURF) and later extended in~\cite{BaroffioRCTT:TIP}. To the best of the authors' knowledge, the encoding of streams of binary features has not been addressed in the previous literature.
Furthermore, the interest of the scientific community in this kind of problem is witnessed by the creation of a new MPEG ad-hoc group, namely Compact Descriptors for Video Analysis (CDVA), which has recently started its activities~\cite{CDVA}. CDVA targets the standardization of the extraction and coding of visual features in application scenarios ranging from video retrieval, automotive, surveillance, industrial monitoring, etc., in which video, rather than images, plays a key role. 

The rest of this paper is organized as follows. Section~\ref{sec:local_setup} states the problem of coding sets of local binary descriptors, defining the properties of the features to be coded, whereas Section~\ref{sec:local_algorithm} illustrates the coding architecture. Section~\ref{sec:global_setup} introduces the problem of coding Bag-of-Visual-Words extracted from a video sequence and Section~\ref{sec:global_algorithm} defines the coding algorithms. Section~\ref{sec:experiments} is devoted to defining the experimental setup and reporting the results. Finally, conclusions are drawn in Section~\ref{sec:conclusions}.

\section{Coding local features: problem statement}\label{sec:local_setup}
Let $\Image_n$ denote the $n$-th frame of a video sequence,
 % of size $N_x \times N_y$, 
which is processed to extract a set of local features $\Descr_n$. First, a keypoint detector is applied to identify a set of interest points. Then, a descriptor is applied on the (rotated) patches surrounding each keypoint. Hence, each element of $\descr_{n,i} \in \Descr_n$ is a visual feature, which consists of two components: i) a 4-dimensional vector $\coord_{n,i} = [x, y, \scale, \orient]^T$, indicating the position $(x,y)$, the scale $\scale$ of the detected keypoint, and the orientation angle $\theta$ of the image patch; ii) a $\Dim$-dimensional \emph{binary} vector $\dv_{n,i} \in \{0,1\}^\Dim$, which represents the descriptor associated to the keypoint $\coord_{n,i}$.

We propose a coding architecture which aims at efficiently coding the sequence $\{\Descr_n\}_{n = 1}^N$ of sets of local features. In particular, we consider both lossless and lossy coding schemes: in the former, the binary description vectors are preserved throughout the coding process, whereas in the latter only a subset of $K < \Dim$ descriptor elements is lossless coded, thus discarding a part of the original data. 
% The latter approach is lossy, since a lossless coding scheme is applied only to a subset of the descriptor elements. 
Each decoded descriptor can be written as $\descrq_{n,i} = \{\coordq_{n,i}, \dq_{n,i}\}$. The number of bits necessary to encode the $M_n$ visual features extracted from frame $\Image_n$ is equal to
\begin{equation}
	\Rate_n = \sum_{i = 1}^{M_n} (\Rate_{n,i}^p + \Rate_{n,i}^d).
\end{equation} 
That is, we consider the rate used to represent both the location of the keypoint, $\Rate_{n,i}^p$, and the descriptor itself, $\Rate_{n,i}^d$. 
For both the lossless and the lossy approach, no distortion is introduced during the coding process in the received descriptor elements. Nonetheless, since in the lossy case part of the descriptor elements are discarded, the accuracy of the visual analysis task might be affected.
% The distortion is measured in terms of the mean square error between the original and decoded descriptor, averaged over the descriptors extracted from $\Image_n$
% \begin{equation}
% 	\Dist_n = \frac{1}{M_n \Dim} \sum_{i = 1}^{M_n} \| \dq_{i,n} - \dv_{i,n} \|_2^2,
% \end{equation} 
% where $\|\cdot\|_2$ denotes the $l$-2 norm. 

As for the component $\coordq_{n,i}$, we decided to encode the coordinates of the keypoint, the scale and the local orientation i.e., $\coordq_{n,i} = [\tilde{x}, \tilde{y}, \tilde{\scale}, \tilde{\orient}]^T$. 
Although some visual analysis tasks might not require this information, it could be used to refine the final results. For example, it is necessary when the matching score between image pairs is computed based on the number of matches that pass the spatial verification step using, e.g., RANSAC~\cite{DBLP:conf/cvpr/PhilbinCISZ07} or weak geometry checking~\cite{DBLP:conf/eccv/JegouDS08}.  Most of the detectors produce floating point values as keypoint coordinates, scale and orientation, thanks to interpolation mechanisms. Nonetheless, we decided to round such values with a quantization step size equal to 1/4 for the coordinates and the scale, and $\pi/16$ for the orientation, which has been found to be sufficient for typical applications~\cite{BaroffioRCTT:ICIP2013, BaroffioRCTT:TIP}. 

% Note that the proposed coding architecture, designed to encode visual features extracted from video sequences, can be straightforwardly adapted also to the context of sets of descriptors extracted from multiple cameras observing the same scene.

\begin{figure*}[t]
	\centering
		\includegraphics[trim=0cm 10.2cm 0cm 1.7cm, clip=true,width=0.65\textwidth]{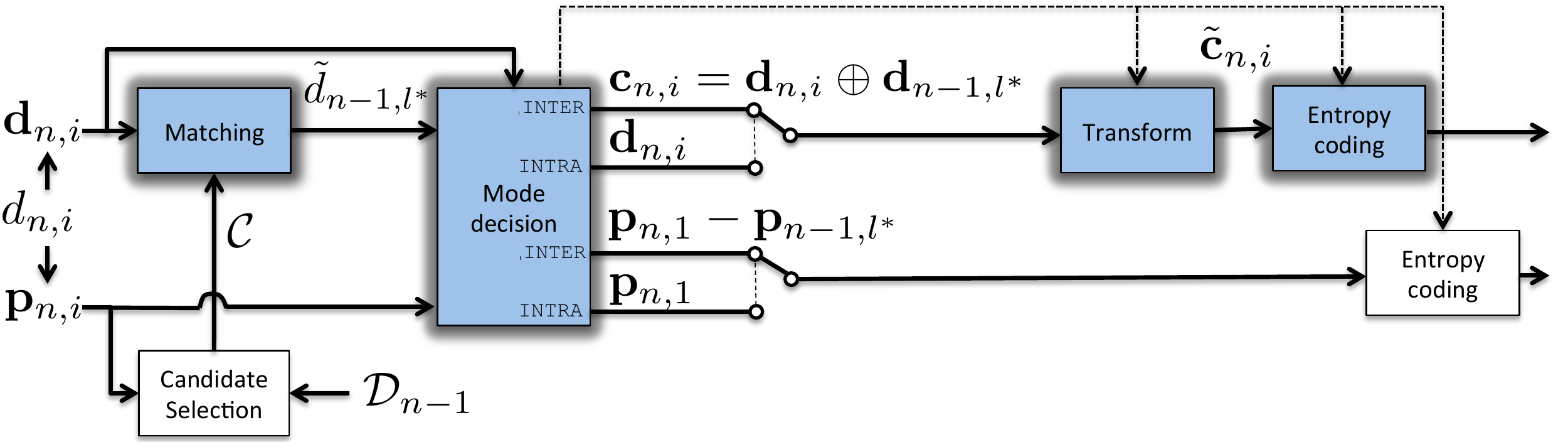}
	\caption{Block diagram of the proposed coding architecture. The highlighted functional modules needed to be revisited due to the binary nature of the source.}
	\label{fig:block_diagram} 
\end{figure*}
%MT aggiungere colore di sfondo blocchi modificati (matching, mode decision, transform, entropy coding del descrittore)
%LB ok
\section{Coding local features: algorithms}\label{sec:local_algorithm}
Figure~\ref{fig:block_diagram} illustrates a block diagram of the proposed coding architecture. The scheme is similar to the one we recently proposed for encoding real-valued visual features~\cite{BaroffioRCTT:ICIP2013, BaroffioRCTT:TIP}. However, we highlighted the functional modules that needed to be revisited due to the binary nature of the source.

\subsection{Intra-frame coding}\label{sec:intra_local}

In the case of intra-frame coding, local features are extracted and encoded separately for each frame. In our previous work we proposed an intra-frame coding approach tailored to binary descriptors extracted from still images~\cite{RedondiBACT:ICIP2013}, which is briefly summarized in the following. In binary descriptors, each element represents the binary outcome of a pairwise comparison. The descriptor elements (dexels) are 
% potentially 
statistically dependent, and it is possible to model the descriptor as a binary source with memory. 

Let $\pi_j$, $j \in [1,\Dim]$ represent the $j$-th element of a binary descriptor $\dv \in \{0,1\}^\Dim$. The entropy of such a dexel can be computed as
\begin{equation}
	H(\pi_{j}) = -p_{j}(0)\log_2(p_{j}(0)) -p_{j}(1)\log_2(p_{j}(1)),
\end{equation}
where $p_j(0)$ and $p_j(1)$ are the probability of $\pi_j = 0$ and $\pi_j = 1$, respectively. Similarly, the conditional entropy of dexel $\pi_{j_1}$ given dexel $\pi_{j_2}$ can be computed as
\begin{equation}
	H(\pi_{j_{1}}|\pi_{j_{2}}) = \sum_{x \in \{0,1\}, y \in \{0,1\}} p_{j_{1},j_{2}}(x,y) \log_{2}\frac{p_{j_{2}}(y)}{p_{j_{1},j_{2}}(x,y)},
\end{equation}

with $j_1, j_2 \in [1,\Dim]$. Let $\tilde{\pi}_{j}$, $j = 1,\dots,\Dim$, denote a permutation of the dexels, indicating the sequential order used to encode a descriptor. The average code length needed to encode a descriptor is lower bounded by
\begin{equation}
	R = \sum_{j=1}^{P}H(\tilde{\pi}_{j}|\tilde{\pi}_{j-1}, \ldots, \tilde{\pi}_{1}).
\end{equation}
In order to maximize the coding efficiency, we aim at finding the permutation of dexels $\tilde{\pi}_{1}, \ldots, \tilde{\pi}_{\Dim}$ that minimizes such a lower bound. For the sake of simplicity, we model the source as a first-order Markov source. That is, we impose $H(\tilde{\pi}_{j}|\tilde{\pi}_{j-1}, \dots \tilde{\pi}_{1}) = H(\tilde{\pi}_{j}|\tilde{\pi}_{j-1})$. Then, we adopt the following greedy strategy to reorder the dexels:
\begin{equation}
	\tilde{\pi}_{j} = \begin{cases} 
						\arg \min_{\pi_{j}} H({\pi_{j}}) & j = 1 \\  
						\arg \min_{\pi_{j}} H({\pi_{j}}|\tilde{\pi}_{j-1}) & j \in [2, \Dim]
					  \end{cases}
\end{equation}
The reordering of the dexel is described by means of a permutation matrix $\T^{\texttt{INTRA}}$, such that $\cq_{n,i} = \T^{\texttt{INTRA}}\dv_{n,i}$. 
Note that such optimal ordering is computed offline, thanks to a training phase, and shared between both the encoder and the decoder. As such, this does not require additional bitrate. 

\subsection{Inter-frame coding}\label{sec:inter_local}
As for inter-frame coding, each set of local features $\Descr_{n}$ is coded resorting to a reference set of features. In this work we consider as a reference the set of features extracted from the previous frame, i.e., $\Descr_{n-1}$. Considering a descriptor $\descr_{n,i}$, $i = 1, \ldots, M_n$, the encoding process consists in the following steps:

\begin{itemize}[-]
	\item \emph{Descriptor matching}: Compute the best matching descriptor in the reference frame, i.e., 
	\begin{equation}\label{eq:RDO_ME}
		\dv_{n-1, l^*} = \text{arg} \min_{l \in \mathcal{C}} D(\dv_{n,i}, \dv_{n-1,l}) + \lambda\Rate_{n,i}^{p, \texttt{INTER}}(l),
	\end{equation}
	where $D(\dv_{n,i}, \dv_{n-1,l}) = {\|\dv_{n,i} - \dv_{n-1,l}\|}_{0}$ is the Hamming distance between the descriptors $\dv_{n,i}$ and $\dv_{n-1,l}$, $\Rate_{n,i}^{p, \texttt{INTER}}(l)$ is the rate needed to encode the keypoint motion vector and $l^*$ is the index of the selected reference feature used in the next steps. We limit the search for a reference feature within a given set $\mathcal{C}$ of candidate features, i.e., the ones whose coordinates and scales are in the neighborhood of $\descr_{n,i}$, in a range of $(\pm\Delta x, \pm\Delta y, \pm \Delta\sigma)$. 
	The prediction residual is computed as $\cv_{n,i} = \dv_{n,i} \oplus \dv_{n-1, l^*}$, that is, the bitwise $XOR$ between $\dv_{n,i}$ and $\dv_{n-1, l^*}$.
	
	\item \emph{Coding mode decision}: Compare the cost of inter-frame coding  with that of intra-frame coding, which can be expressed as
	\begin{equation}\label{eq:RDO_INTRA}
				J^{\texttt{INTRA}}(\descr_{n,i}) = \Rate_{n,i}^{p, \texttt{INTRA}} + \Rate_{n,i}^{d, \texttt{INTRA}},
    \end{equation}
    \begin{equation}
	J^{\texttt{INTER}}(\descr_{n,i}, \descrq_{n-1,l^*}) =  \Rate_{n,i}^{p, \texttt{INTER}}(l^*) + \Rate_{n,i}^{d, \texttt{INTER}}(l^*),
	\end{equation}
	where $\Rate_{n,i}^{p}$ and $\Rate_{n,i}^{d}$ represent the bitrate needed to encode the location component (either the location itself or location displacement) and the one needed to encode the descriptor component (either the descriptor itself or the prediction residual), respectively.
	If $J^{\texttt{INTER}}(\dv_{n,i}, \dv_{n-1,l^*}) < J^{\texttt{INTRA}}(\dv_{n,i})$, then inter-frame coding is the selected mode. Otherwise, proceed with intra-frame coding.
	
	\item \emph{Intra-descriptor transform}: This step aims at exploiting the spatial correlation between the dexels. If intra-frame is the selected coding mode, then the dexels of $\dv_{n,i}$ are reordered according to the permutation algorithm presented in Section~\ref{sec:intra_local}. Similarly, a reordering strategy can be applied also in the case of inter-frame coding, in this case considering the prediction residual $\cv_{n,i}$, that is, $\cq_{n,i} = \T^{\texttt{INTER}}\cv_{n,i}$. Note that, in general, $\T^{\texttt{INTER}} \ne  \T^{\texttt{INTRA}}$

	\item \emph{Entropy coding}: Finally, the sets of local features are entropy coded. In the case of intra-frame coding, for each local feature, it is necessary to encode the reordered descriptor and the quantized location component. Otherwise, for inter-frame coding, it is necessary to encode: i) the identifier of the matching keypoint in the reference frame and the displacement in terms of position, scale and orientation of the keypoint with respect to the reference, which require $\Rate_{n,i}^{p, \texttt{INTER}}(l^*)$ bits; ii) the reordered prediction residual $\cq_{n,i}$. 
	
	For both intra-frame and inter-frame coding, the probabilities of the symbols (respectively, descriptor elements or prediction residuals) used for entropy coding are learned from a training set of frames. In particular, for each of the $\Dim$ dexels, we estimated the conditional probability of each symbol, given the previous one defined by the optimal permutation. 
	% Such a procedure is applied also in the case of inter-frame coding, exploiting a training set of prediction residuals. 
	The estimated probabilities are then exploited to entropy code the features.
\end{itemize}

\subsection{Descriptor element selection}\label{sec:BAMBOO}

The lossless coding architecture described in the previous section can be used to encode all the $\Dim$ elements of the original binary descriptor. However, in order to operate at lower bitrates, it is possible to decide to code only a subset of $K < \Dim$ descriptor elements. In our previous work we explored different methods that define how to select the dexels to be retained~\cite{RedondiBACT:ICIP2013, BAMBOO_MMSP, BAMBOO_ICASSP}. In this work, we employed the greedy asymmetric pairwise boosting algorithm described in~\cite{BAMBOO_ICASSP} in order to iteratively select the most discriminative descriptor elements. To this end, we used a training set of image patches~\cite{bath26111}, along with the ground truth information defining whether two image patches refer to the same physical entity. At each step, the asymmetric pairwise boosting algorithm selects the dexel that minimizes a cost function, which captures the error resulting from the wrong classification of matching and non-matching patches. The output of this procedure is a set of dexels, ordered according to their discriminability. Hence, given a target descriptor size $K < \Dim$, it is possible to encode only the first $K$ descriptor elements selected by this algorithm.

\section{Global descriptors based on binary visual features}\label{sec:global_setup}
Let $\Image_n$ denote the $n$-th frame of a video sequence, which is processed to extract a set of local features $\Descr_n$. A global representation for the frame $\Image_n$ can be computed, starting from such set of local image descriptors. The key idea behind the Bag-of-Visual-Words (BoVW) approach is to quantize each local feature into one visual word. To this end, a vocabulary $\Dict = \{\vw_1, \dots, \vw_\numvw\}$ composed of $\numvw$ %LB K already refers to target descriptor size
 distinct visual words has to be computed. Traditional approaches to the creation of BoVW models are based on real-valued local descriptors such as SIFT~\cite{DBLP:journals/ijcv/Lowe04} or SURF~\cite{DBLP:conf/eccv/BayTG06}. In this context, a large set of descriptors $\dv \in \mathbb{R}^K$ is exploited for learning the vocabulary, along with a clustering algorithm such as $k$-means (with $k = \numvw$) based on Euclidean distance~\cite{DBLP:conf/iccv/SivicZ03}, Gaussian Mixture Model~\cite{Fern11supervisedlearning}, etc.

More recently, the problem of constructing a BoVW model starting from sets of binary local descriptors was addressed in~\cite{Galvez_BOBVW}. Analogously to the case of real-valued descriptors, a dictionary is learned starting from a large set of descriptors $\dv \in \{0, 1\}^K$. To this end, a naive approach would consist in k-means clustering paired with Euclidean distance~\cite{Paratte}. Besides, clustering techniques tailored to the peculiar nature of the signal at hand have been introduced. In particular, $k$-medoids and $k$-medians algorithms, paired with Hamming distance, have been successfully exploited for creating the dictionary~\cite{Galvez_BOBVW}.
%MT e noi cosa usiamo? qui si parla di metodi ad hoc per binary features, mentre negli esperimenti si menziona k-means

Then, given a vocabulary that consists of $\numvw$ of visual words $\Dict = \{\vw_1, \dots, \vw_\numvw\}$ learned offline and a set of visual features $\Descr_n$ extracted from the frame $\Image_n$, a global descriptor is obtained by mapping such a set of features to a fixed-dimensional vector $\bag_n \in \mathbb{R}^\numvw$. The simplest strategy is to apply hard quantization, which assigns each feature $\dv \in \Descr_n$ to the nearest visual word's centroid $\vw_j \in \Dict$. The resulting global descriptor $\bag_n = [v_1, \dots, v_\numvw]^T$ is a histogram, where $v_j$ represents the number of local features in $\Descr_n$ having the dictionary word $\vw_j$ as nearest neighbor. Soft quantization represents a more sophisticated alternative to hard quantization, mapping each local feature to multiple visual words. Finally, % several normalization techniques to be applied on the global descriptor 
several techniques for normalizing the global descriptor $\bag_n$ have been proposed, aimed at improving matching performance. A widely accepted approach consists in adopting the tf-idf weighting scheme, followed by $L_2$ normalization~\cite{Philbin2007}. The former gives more emphasis to rare visual words and less importance to common ones, whereas the latter avoids short vectors, i.e. BoVWs built starting from few local features, to be penalized during the matching stage.

\section{Coding global features}\label{sec:global_algorithm}
For each frame $\Image_n$ of an input video sequence, a set of binary local features $\Descr_{n}$ is extracted and mapped to a $\numvw$-dimensional global descriptor $\bag_n = [v_1, \dots, v_\numvw]^T$ by applying the procedure described in Section~\ref{sec:global_setup}. We propose a coding architecture which aims at effectively encoding the sequence $\{\bag_n\}_{n=1}^N$ of global image descriptors. In particular, such a lossy coding architecture enables the decoder to reconstruct an approximation $\{\recbag_n\}_{n=1}^N$ of the original sequence of global descriptors.

Differently from the case of local descriptors, the coordinates of the keypoints are disregarded during the construction of the BoVW and they are not encoded. Hence, the number of bits needed to encode a Bag-of-Visual-Words $\bag_n$ extracted from frame $\Image_n$ is equal to $R_n = \sum_{j=1}^{\numvw}{R_{n,j}^\bag}$, i.e., the sum of the number of bits needed to encode each component of the vector $\bag_n$.

\subsection{Intra-frame coding}\label{sec:intra_bow}
The Intra-frame coding approach is based on a frame-by-frame processing scheme, in which the global descriptor extracted from the frame $\Image_n$ is encoded independently from the ones extracted from other frames. Considering a baseline architecture, uniform scalar quantization with step size $\Delta_j$ is applied to each element $v_{n,j}, j = 1, \dots, \numvw$ of the global descriptor $\bag_n$, that is
\begin{equation}\label{eq:bow_quant}
	\qbagdex_{n,j} = \left \lfloor \frac{v_{n,j}}{\Delta_j} \right \rfloor.
\end{equation}
Since the vectors are normalized according to a tf-idf weighting scheme, the same quantization step size $\Delta_j = \Delta, \; j = 1, \dots, \numvw$, is fixed for each visual word. 

The quantization index $\qbagdex_{n,j}$ is considered as the outcome of a discrete memoryless source and entropy coded. To this end, the probabilities of the quantization symbols are estimated offline and fed to an arithmetic coder, so that the corresponding rate is equal to $R_{n,j}^\bag = -\log_2 (p(\qbagdex_{n,j}))$.

% 
% $p(q_{j} = X_p),\;j = 1, \dots, M,$ where $X_p,\; p = 1, \dots, S$ are the possible values for the quantization symbol $q_j$, are estimated offline and fed to the entropy coder. 
% 
% Then, the elements of the vector are entropy coded using $R_n$ bits. The probabilities of the quantization symbols $p(q_{j} = X_p),\;j = 1, \dots, M,$ where $X_p,\; p = 1, \dots, S$ are the possible values for the quantization symbol $q_j$, are estimated offline and fed to the entropy coder. 
% Considering the BoVW $\bag_n$, the expected number of bits required to encode such a signal is equal to
% \begin{equation}
% 	R_n = \sum_{j=1}^{M}{R_{n,j}^\bag} = \sum_{j=1}^{M} \log_2 (p(q_{j} = \qbagdex_{n, j})).
% \end{equation}

\subsection{Inter-frame coding}\label{sec:inter_bow}

In the case of inter-frame coding, local features are extracted on a frame-by-frame basis and quantized into BoVWs in order to obtain a sequence of global descriptors $\{\bag_n\}_{n=1}^N$. Then, differently from intra-frame coding, temporal redundancy is exploited in the coding phase: the global descriptor $\bag_n$ extracted from frame $\Image_n$ is encoded using $\bag_{n-1}$ as reference.
% , that is, the BoVW extracted from the previous frame $\Image_{n-1}$. 
In particular, each descriptor element $v_{n,j}$ is encoded having $v_{n-1,j}$ as context. 

To this end, considering a quantization step size $\Delta_j$, the quantization symbols $\qbagdex_{n,j}$ are obtained according to Equation~\eqref{eq:bow_quant}, and then entropy coded using $R_n$ bits. Similarly to the case of intra-frame coding, the statistics of the quantization symbols are estimated at training time. In particular, given a sufficiently large sequence of training global descriptors, a training phase aims at estimating the probabilities $p(q_{n, j} = X_p | q_{n - 1, j} = X_q)$, i.e. the probabilities of the $j$-th dexel at frame $\Image_n$ assuming value $X_p$, given the $j$-th dexel at frame $\Image_{n - 1}$ having value $X_q$. 
%
%At test time, given a global descriptor $\bag_{n}$, the corresponding quantization symbols are obtained by applying scalar quantization and fed to an entropy coder. 
An arithmetic coder is used to entropy code the quantization symbols, with an expected number of bits that amounts to

\begin{equation}
	R_n = \sum_{j=1}^{M}{R_{n,j}^\bag} = \sum_{j=1}^{M} \log_2 (p(q_{n, j} = \qbagdex_{n, j}|q_{n-1, j} = \qbagdex_{n-1, j})).
\end{equation}
\section{Experiments}\label{sec:experiments}
We validated the effectiveness of the feature coding architectures and compared the two different paradigms, namely ``Analyze-Then-Compress'' (ATC) and ``Compress-Then-Analyze'' (CTA), on two traditional visual analysis tasks:
\begin{itemize}
	\item \emph{Homography estimation}. Several high- and low-level visual analysis tasks, including camera calibration, 3D reconstruction, structure-from-motion, tracking, etc. might require the estimation of the homography defining the geometrical relationship between two frames with homogeneous visual content. In this scenario, local features can be conveniently used to find correspondences between pixel locations in different frames or views. Conversely, global features based on BoVW do not represent a viable option, since they do not include any geometrical information about the visual content.
	\item \emph{Content Based Retrieval} (CBR). Content Based Retrieval is a traditional, yet challenging, task within the computer vision community. Given an input query in the form of some kind of visual content, the goal is to retrieve the relevant multimedia documents within a large database. Accuracy and computational efficiency are key tenets to be considered when implementing algorithms for CBR, which typically target large scale scenarios. Our test considers an input query in the form of a video clip, with the goal of retrieving the most relevant database images. In this scenario, both global and local features are considers, in order to explore a trade-off between accuracy and computational efficiency. 
\end{itemize}

\subsection{Data sets}
\subsubsection{Training data sets} The methods discussed for encoding binary local descriptors require the knowledge of the probabilities of the symbols to operate the entropy coder, which were estimated from training sequences. To this end, we employed three video sequences, namely \emph{Mother}, \emph{News} and \emph{Paris}~\cite{DeSimoneTNTE:ICASSP2010}. The training video sequences were also exploited to obtain the optimal coding order of dexels for both intra- and inter-frame coding, as illustrated in Section~\ref{sec:local_algorithm}. Furthermore, a dataset of patches~\cite{bath26111} was exploited along with an asymmetric pairwise boosting algorithm~\cite{BAMBOO_ICASSP}~\cite{BAMBOO_MMSP} in order to identify the $K$ most discriminative dexels according to the method presented in Section~\ref{sec:BAMBOO}.

	In the case of BoVW-based global descriptors, the visual word dictionary was estimated exploiting a large database of images, namely \emph{VOC2010}~\cite{pascal-voc-2010}, whereas the statistics of the coding symbols for both intra- and inter-frame coding architectures were estimated offline, resorting to a sufficiently long video sequence, namely \emph{Rome in a nutshell}, which consists of 15375 frames.
%MT add number of frames
%LB ok
%MT Add footnote with link for download, if this is not a standard sequence
%LB TODO, check copyright

\subsubsection{Test data sets}
\begin{figure}[]
		\centering
		\subfigure[]{\includegraphics[trim=0cm 2.5cm 15cm 0cm, clip=true,width=0.35\textwidth]{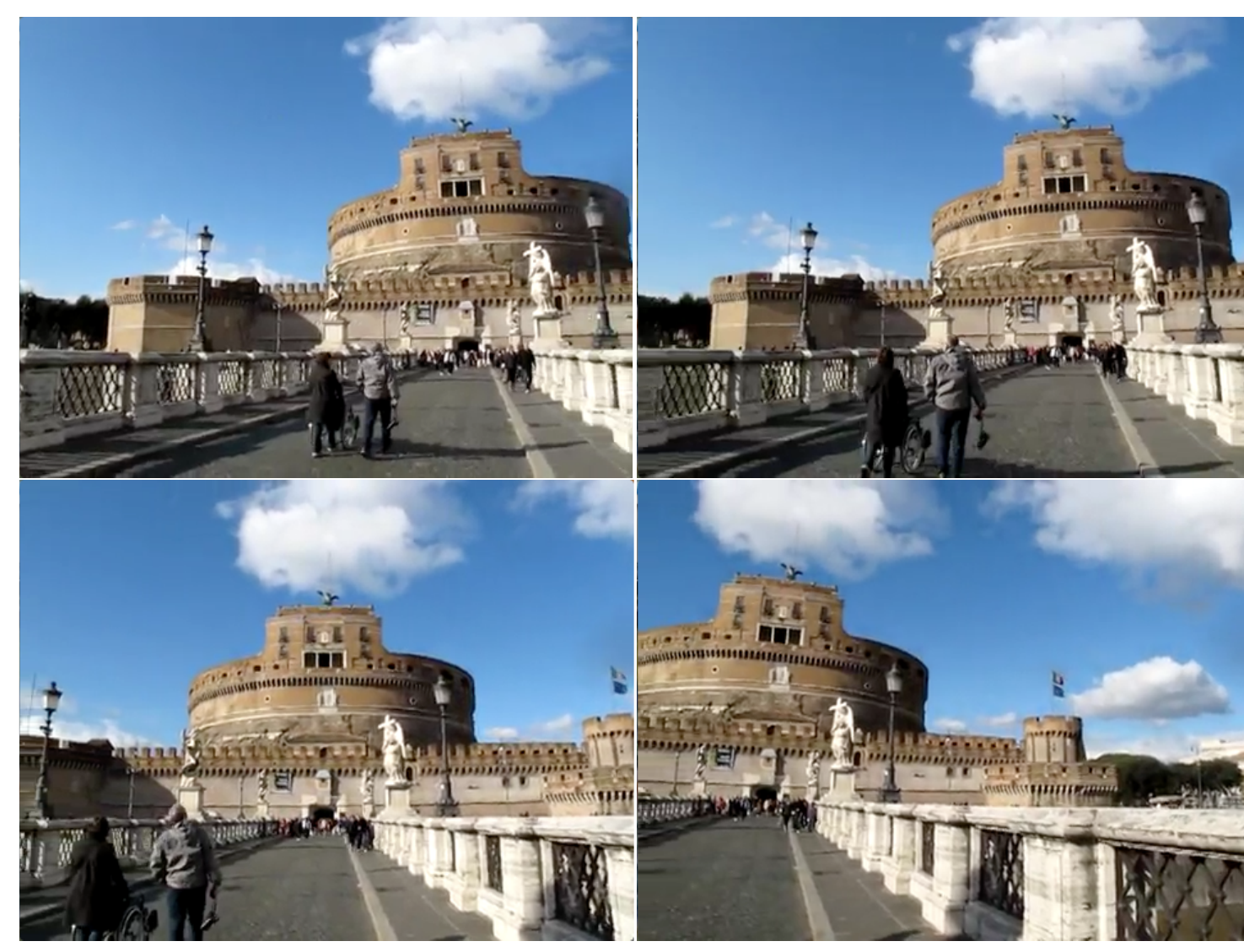}\label{fig:RD_foreman}}
		\subfigure[]{\includegraphics[trim=2cm 8.1cm 2cm 8cm, clip=true,width=0.30\textwidth]{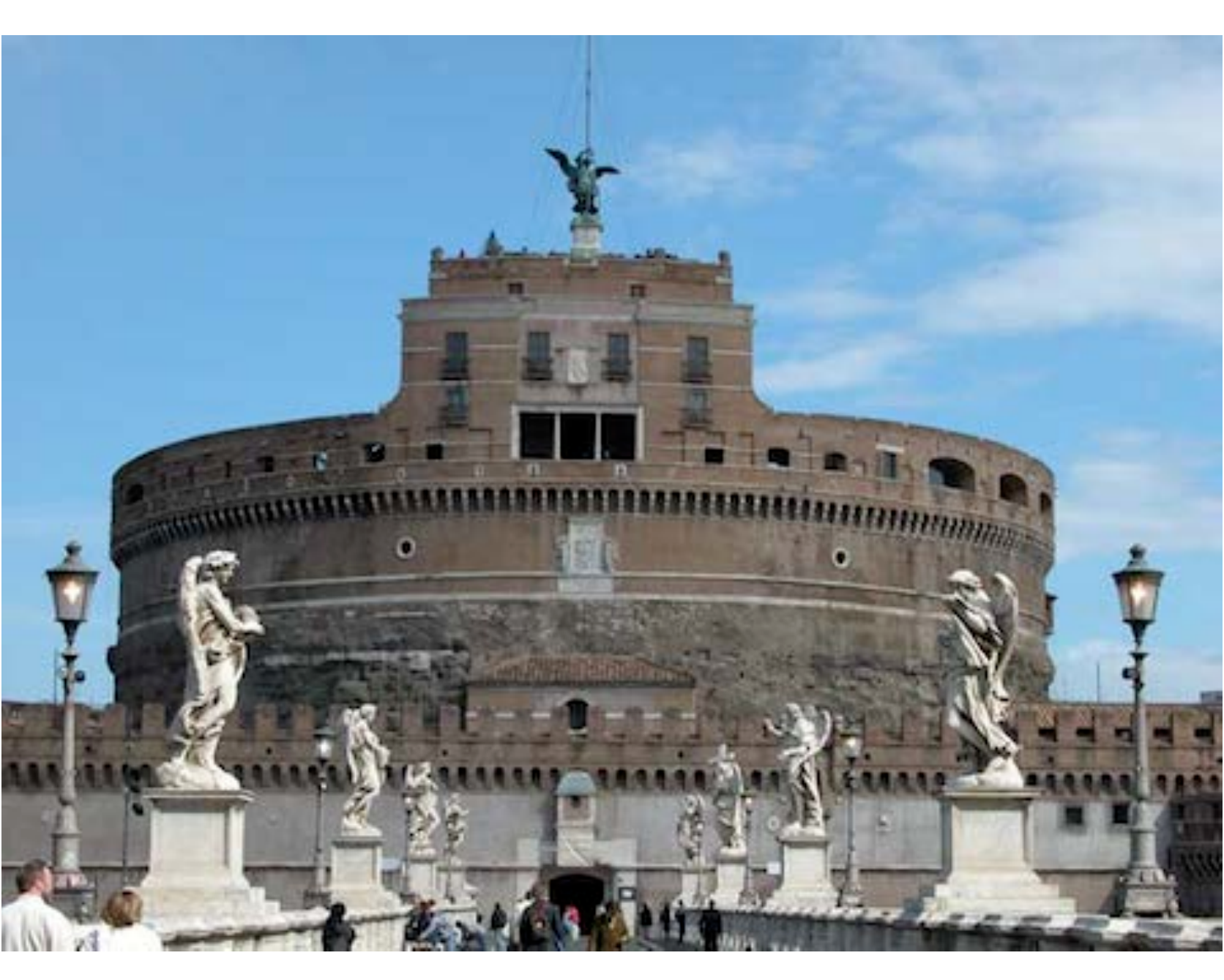}\label{fig:RD_ada_foreman}}
		\caption{a) Four frames sampled from one of the query videos employed for the retrieval task. b) A matching database image.}
		\label{fig:retrieval_video} 
	\end{figure}

	First, the coding architecture was evaluated on three video sequences, namely \emph{Hall}, \emph{Mobile} and \emph{Foreman}, to investigate the bitrate saving which can be obtained by properly encoding the binary features. Then, for the \emph{Homography Estimation} test, we used a publicly available dataset for visual tracking~\cite{Gauglitz_IJCV2011}, consisting in a set of video sequences, each containing a planar texture subject to a given motion path. For each frame of each sequence, the homography that warps such frame to the reference is provided as ground truth. The sequences have a resolution of 640 $\times$ 480 pixels at 15 fps and a length of 500 frames (33.3 seconds). Finally, for the \emph{Content Based Retrieval} (CBR) test, a set of 10 query video sequences was used, each capturing a different landmark in the city of Rome with a camera embedded in a mobile device~\cite{romelandmark}. The frame rate of such sequences is equal to 24fps, whereas the resolution ranges from 480x360 pixels (4:3) to 640x360 pixels (16:9). The first 50 frames of each video were used as query. On average, each query video corresponds to 9 relevant images representing the same physical object under different conditions and with heterogeneous qualities and resolutions. Then, distractor images randomly sampled from the \emph{MIRFLICKR-1M} dataset, so that the final database contains 10k images. As an example, Figure~\ref{fig:retrieval_video} shows some frames of a query sequence, along with a relevant image to be retrieved. The dataset is publicly available for download at XX.
	%LB make dataset publicly available?
	%MT yes
	%LB check copyright issues
	
	\subsection{Methods}
	\subsubsection{ATC-Training}

		The proposed coding architecture can be applied to any kind of local binary feature. Hence, in our experiments we evaluated the use of two different state-of-the-art binary features, namely BRISK~\cite{DBLP:conf/iccv/LeuteneggerCS11} and BINBOOST~\cite{Trzcinski13a}. The detection threshold was set equal to 70 for both BRISK and BINBOOST. All other parameters were left equal to their default values. In both cases, the parameters $x$, $y$ and $\sigma$, representing the location and the scale of each keypoint, were rounded to the nearest quarter of unit.  Descriptors consisting of $P = 512$ and $P = 256$ dexels for BRISK and BINBOOST, respectively, were extracted from the training video sequences, using the original implementations of the feature extraction algorithms provided by authors. 

		As for BRISK, we considered a set of target descriptor sizes $K = \{512, 256, 128, 64, 32, 16, 8\}$. For each size, we employed the dexel selection algorithm presented in Section~\ref{sec:BAMBOO} in order to identify the elements to be retained. Then, in the case of intra-frame coding and for each descriptor length, the optimal coding order and the corresponding coding probabilities were estimated according to the procedure introduced in Section~\ref{sec:intra_local}. Instead, in the case of inter-frame coding, for each descriptor a match was found within the features extracted from the previous frame, according to the method presented in Section~\ref{sec:inter_local}. Similarly to the case of intra-frame coding, a coding-wise optimal permutation of the elements of the binary prediction residual was computed, and the corresponding coding probabilities were estimated.
		As to BINBOOST, we considered a set of target descriptor sizes $K = \{256, 128, 64, 32, 16, 8\}$. For each size, the first $K$ dexels of the BINBOOST descriptor were retained. Then, similarly to the case of BRISK, coding-wise optimal dexel permutations for intra- and inter-frame coding were computed, along with the probabilities of the coding symbols. 

		In the case of BoVW-based global descriptors, we fixed a set of target sizes for the dictionary of visual words $M = \{1024, 4096, 16384\}$. Then, for each possible dictionary size, BRISK or BINBOOST descriptors were extracted from the training set of images and vector quantization was applied in order to identify $M$ visual words. Both k-means and k-medians algorithms have been tested for the dictionary construction stage, yielding similar results in terms of rate-accuracy performance. Furthermore, global descriptors based on BRISK and BINBOOST local features achieve very similar results. In the following, we refer to the best performing setup, that is, k-means clustering applied to BRISK descriptors, initialized according to the k-means++~\cite{Arthur:2007:KAC:1283383.1283494} algorithm, and Euclidean distance. The output of this first stage is a dictionary composed of $M$ visual words each represented by a $P$-dimensional vector, where $P = 512$ ($P = 256$) is the size of BRISK descriptors (BINBOOST descriptors). Then, a training video sequence was adopted to compute the coding probabilities. For each frame, local features were extracted and the global descriptor was computed by hard assigning each feature to its nearest neighbor within the dictionary, according to the procedure presented in Section~\ref{sec:global_setup}. Then, for each target quantization step size $\Delta = \{0.01, 0.05, 0.1, 0.2\}$, global descriptors were quantized and the coding probabilities for both intra- and inter-frame were computed according to the algorithms introduced in Section~\ref{sec:global_algorithm}.
		%MT abbiamo fatto esperimenti con BoVW e BINBOOST?
		%LB TODO

		Concerning the CBR test, a representation of each database image had to be computed, in the form of both a set of local features and a global descriptor. To this purpose,
		% , considering~\ATC, 
		for each image a set of local features was extracted and stored. Furthermore, such a set was also exploited to compute a BoVW-based global descriptor that was stored, too. % As for~\CTA, a similar procedure was performed resorting to SIFT features and both local and global image representations are stored. 

	\subsubsection{CTA-Training}

		The \emph{Compress-Then-Analyze} paradigm relies on traditional video compression, paired with state-of-the-art visual features extraction algorithms. In the case of local features, no training was needed. Instead, in the case of global features, a dictionary of visual words had to be learned in order to compute the BoVW representation of an image. The dimensionality of the dictionary was fixed to $M = 16384$ visual words and, similarly to the case of ATC paradigm, SIFT local features were extracted from the \emph{VOC2010} dataset and clustered into $M$ visual words, once again resorting to k-means (k-means++ initialization). 

	\subsubsection{ATC-Testing}

		within the ATC paradigm, we distinguished between several different schemes:
		\begin{itemize}
			\item \texttt{BRISK/BINBOOST - INTRA}: all binary local features (either BRISK or BINBOOST) were encoded resorting to an intra-frame coding scheme.
			\item \texttt{BRISK/BINBOOST - INTER}: all binary local features were encoded resorting to an inter-frame coding scheme.
			\item \texttt{BRISK/BINBOOST - INTRA/INTER}: for each binary local feature, a 2-way coding mode decision module was used to select the best coding mode between INTRA and INTER.
			\item \texttt{BoVW - INTRA}: all global features were encoded resorting to an intra-frame coding scheme.
			\item \texttt{BoVW - INTER}: all global features were encoded resorting to an inter-frame coding scheme.
		\end{itemize}
		%MT come mai qui usiamo BOW anziché BoVW come nel resto del paper?
		%LB modificato qui e sulle figure

	\subsubsection{CTA-Testing}

		within the CTA paradigm, we distinguished between two different schemes:
		\begin{itemize}
			\item \texttt{SIFT - INTER}: visual content was encoded resorting to H.264/AVC coder, SIFT features were employed.
			\item \texttt{BoVW - INTER}: visual content was encoded resorting to H.264/AVC coder, BoVW-based global features were employed.

		\end{itemize}

		For the tests to be as fair as possible, the video coding scheme and the visual feature coding scheme were configured to operate under comparable conditions. In particular, the following settings were employed with the x264 library, by adopting coding tools that are supported by the H.264/AVC baseline profile, which is tailored for wireless communications:
		\begin{itemize}
			\item number of reference frames: 1 (\texttt{--ref 1})
			\item B-frames disabled (\texttt{--bframes 0})
			\item subpixel motion estimation complexity: quarter of pixel (\texttt{--subme 4})
			\item Trellis quantization disabled (\texttt{--trellis 0})
			\item Context-Adaptive Binary Arithmetic Coding (CABAC) disabled (\texttt{--no-cabac})
		\end{itemize}

		The Constant Rate Factor parameter (\texttt{--crf <integer>}) was employed to control the output bitrate. It is important to emphasize that the H.264/AVC standard is the result of many years of optimization, while coding of visual features has only been recently explored. Therefore, some of the coding tools successfully adopted in H.264/AVC (e.g., B-frame, multiple reference frames, etc.), might also be integrated into our coding architecture. This is left to future investigation.

	\subsection{Experiments and evaluation metrics}
		Each visual analysis task was evaluated according to an ad-hoc metric:

		\subsubsection{Homography estimation}
			In the case of $\ATC$, the sets of features $\Descr_n$ were extracted starting from the test sequences. Such sets were filtered, removing the keypoints that did not belong to the planar texture identified by the available ground truth. For each value of the quantization step size $\Delta$, the sets $\Descrt_{n,\Delta}$ were obtained following the $\ATC$ paradigm. For each pair of consecutive frames $\mathcal{I}_n$ and $\mathcal{I}_m$, a homography $\tilde{H}_{nm,\ATC, \Delta}$ was estimated based on $\Descrt_{n,\Delta}$ and $\Descrt_{m, \Delta}$. To this end, the matches between the two sets of features were identified and given as input to the RANSAC algorithm \cite{Fischler:1981:RSC:358669.358692}. 

		As for \CTA, the test sequences were encoded with each one of the quality factors $Q = \{5, 10, \dots, 45\}$. For each frame $\mathcal{I}_n$ of the encoded sequence the sets of features $\Descrt_{n,Q}$ were extracted. Similarly to the $\ATC$ case, the sets of visual features were filtered and for each pair of consecutive frames $\mathcal{I}_n$ and $\mathcal{I}_m$, a homography $\tilde{H}_{nm,\CTA, Q}$ was estimated resorting to $\Descrt_{n,Q}$ and $\Descrt_{m,Q}$.

		The performance of $\ATC$ and $\CTA$ was evaluated in terms of rate-efficiency curves. For the task at hand, efficiency was measured computing the \textit{homography estimation precision}, which was adopted in our previous work~\cite{BaroffioRCTT:TIP} and briefly summarized here for completeness. 
	%MT add ref	
	%LB ok
		Specifically, let $\tilde{H}_{nm}$ denote the homography estimated according to the procedure presented above, following either the $\ATC$ or the $\CTA$ approach. The coordinates of the four corners of the texture $c_{1,n}$, $c_{2,n}$, $c_{3,n}$, $c_{4,n}$ in frame $\Image_n$ were provided as ground truth. Applying the homography $\tilde{H}_{nm}$ to such points, it was possible to estimate the coordinates $\tilde{c}_{1,m}$, $\tilde{c}_{2,m}$, $\tilde{c}_{3,m}$, $\tilde{c}_{4,m}$ in frame $\Image_m$ and compare them with the real coordinates of the corners $c_{1,m}$, $c_{2,m}$, $c_{3,m}$, $c_{4,m}$, also available as ground truth. The \emph{backprojection error} for the frame $\Image_m$ is defined as $\mathcal{E}_{bp}(m) = \frac{1}{4} \sum\limits_{p=1}^4 |\tilde{c}_{p,m} - c_{p,m}|$. An estimated homography was deemed correct if the relative backprojection error was lower than $\epsilon_{bp} = 3$ pixels. Finally, the \emph{homography estimation precision} is defined as the ratio between the number of correctly estimated homographies and the total number of frames.

		\subsubsection{Content Based Retrieval}\label{sec:methods_CBR}
		Considering~\ATC~and given a query video sequence, a set of local features was extracted from each frame of the clip and mapped to a BoVW-based global descriptor, according to the procedure described in Section~\ref{sec:global_setup}. The goal of the task is the retrieval of relevant images within a database consisting of $Z$ elements. Considering traditional applications of CBR, database dimensionality $Z$ ranges from thousands to millions. Hence, matching based on sets of local features might represent an inefficient, or even unfeasible, approach. On the other hand, global image descriptors represent an effective yet computationally efficient solution. Indeed, a two-step approach was proposed~\cite{DBLP:conf/iccv/SivicZ03}, which consist in i) retrieving the top-$k$ relevant results within the database exploiting global descriptors and ii) refine the results of the previous step exploiting local features. Such an approach represents a good tradeoff between task accuracy and computational efficiency, since fast matching based on global features is exploited in order to identify a subset of possibly relevant documents, whereas an accurate re-ranking is performed on such a small subset of data, resorting to local visual features.

		Considering the first stage of the pipeline, i.e. the retrieval of top-$k$ relevant items, the global descriptor extracted from each frame of each test video sequence was matched against the global descriptors of all the database images. Due to the adoption of the weighting and normalization procedure described in Section~\ref{sec:global_setup}, Euclidean distance was employed to compare pairs of global descriptors. Then, database images were ranked according to their distance with respect to the query, in increasing order. The top-$k$ elements of the ranking are the matching candidates for the query at hand. For such a test, we fixed $k = 200$, so that re-ranking is performed only on $2\%$ of database images.
		We evaluated the performance in terms of rate-efficiency curves. In particular, the accuracy of the task was evaluated according to the \emph{Mean Average Precision} (MAP). Given an input query sequence $q$, for each frame $\Image_{q,n}$ it is possible to define the \emph{Average Precision} as 
		\begin{equation}
			AP_{q,n} = \frac{\sum_{k=1}^Z P_{q,n}(k)r_{q,n}(k)}{R_{q,n}},
		\end{equation}
		where $P_{q,n}(k)$ is the precision (i.e., the fraction of relevant documents retrieved) considering the top-$k$ results in the ranked list of database images; $r_{q,n}(k)$ is an indicator function, which is equal to 1 if the item at rank $k$ is relevant for the query, and zero otherwise; $R_{q,n}$ is the total number of relevant document for frame $\Image_{q,n}$ of the query sequence $q$ and $Z$ is the total number of documents in the list.
		The overall accuracy for the query sequence $q$ is evaluated according to

		\begin{equation}
			AP_q = \frac{\sum_{n = 1}^N AP_{q,n}}{N}, 
		\end{equation}

	where $N$ is the total number of frames of the query video $q$. 

	Finally, the \emph{Mean Average Precision} for the CBR task is obtained as
	\begin{equation}\label{eq:MAP}
		MAP = \frac{\sum_{q = 1}^Q AP_q}{Q}, 
	\end{equation}
	that is, the mean of the $MAP_q$ measure over all the query sequences.

	We also considered an alternative way of aggregating the results of a video query $q$, resorting to \emph{Median Rank Aggregation} (MRA). To this end, considering a test sequence of $N$ frames, the retrieval pipeline is executed on each frame $\Image_{q,n}$ leading to $N$ ranked lists of retrieved documents $\mathcal{R}_{q,n}$, $n = 1,\dots, N$. 
	% Considering such an ordered list and for each test frame $\Image_n,\;n=1,\dots,N$, 
	Each database image $D_k,\;k=1,\dots,Z$, can be assigned with a ranking value $\mathcal{P}_{q,n,k}$, equal to its position in the list $\mathcal{R}_{q,n}$. Then, for a database image $D_l$, it is possible to define a relevance score $\mathcal{P}_{q,k}$ to the query $q$ by aggregating the ranking values $\mathcal{P}_{q,n,k}$ obtained for each query frame $\Image_n$. In details, $\mathcal{P}_{q,k}$ is equal to the median value within the set of ranking values $\mathcal{P}_{q,n,k},\;n=1, \dots, N$. Finally, an overall ranking of database images is obtained for a given test sequence, by sorting such documents according to their scores $\mathcal{P}_{q,k}$, in ascending order. Starting from such a ranking, it is possible to compute the \emph{Average Precision} for the query $q$ as
	\begin{equation}
		AP_{q, MRA} = \frac{\sum_{k=1}^Z P_{q,MRA}(k)r_{q,MRA}(k)}{R_{q}},
	\end{equation}
	where $P_{q,MRA}(k)$ is the precision (i.e., the fraction of relevant documents retrieved) considering the top-$k$ results in the ranked list of database images obtained exploiting \emph{Median Rank Aggregation}; $r_{q,MRA}(k)$ is an indicator function, which is equal to 1 if the item at rank $k$ is relevant for the query, and zero otherwise; $R_{q}$ is the total number of relevant document for the query at hand. Finally, the overall MAP is computed as the mean of $AP_{q,MRA}$ over all the query video sequences $q$.

	With respect to the second stage of the CBR task, considering a query video sequence $q$, a set of visual features was extracted from each frame $\Image_{q,n},\;n=1,\dots,N$. Then, such a set of local features was matched against the sets corresponding to the top-$k$ candidate database images identified by means of the procedure detailed above resorting to global features. First, each feature extracted from the query frame was matched with its nearest neighbor in the test set, resorting to Hamming distance or Euclidean distance in the case of $\ATC$ - BRISK  or $\CTA$ - SIFT, respectively. Second, matches were filtered resorting to the ratio test~\cite{DBLP:journals/ijcv/Lowe04}, with ratio parameter set to 0.7. Then, database images were ranked according to the number of matches with the query frame that passed the ratio test. Finally, we computed the MAP metric based on the ranking induced by the number of matches, resorting to the procedure adopted for global descriptors. Similarly, we also obtained results when \emph{Median Rank Aggregation} was used.

	% \hl{
	As a further experiment, we evaluate the effect of temporal subsampling on the overall efficiency of the retrieval pipeline. We tested several different values for the GOP size parameter. When the GOP size is equal to $f$ frames, a global descriptor (set of local descriptors) is sent every $f$ frames. 
	% In details, we tested the configurations GOP = $\{1,2,5,10,25,50\}$. 
	% } 
	With respect to global descriptors, we tested different approaches:

	\begin{itemize}
		\item BoVW-SKIP: considering a Group Of Pictures, a global feature is extracted considering only the first frame of such GOP.
		\item BoVW-GOP: considering a Group of Pictures, global features are extracted from each frame of the GOP, then, the median global descriptor vector is computed and used for the retrieval.
	\end{itemize}

	\subsection{Results}

	\begin{figure*}[t]
		\centering
		\subfigure[]{\includegraphics[trim=0.7cm 0.3cm 1.5cm 0.1cm, clip=true,width=0.42\textwidth]{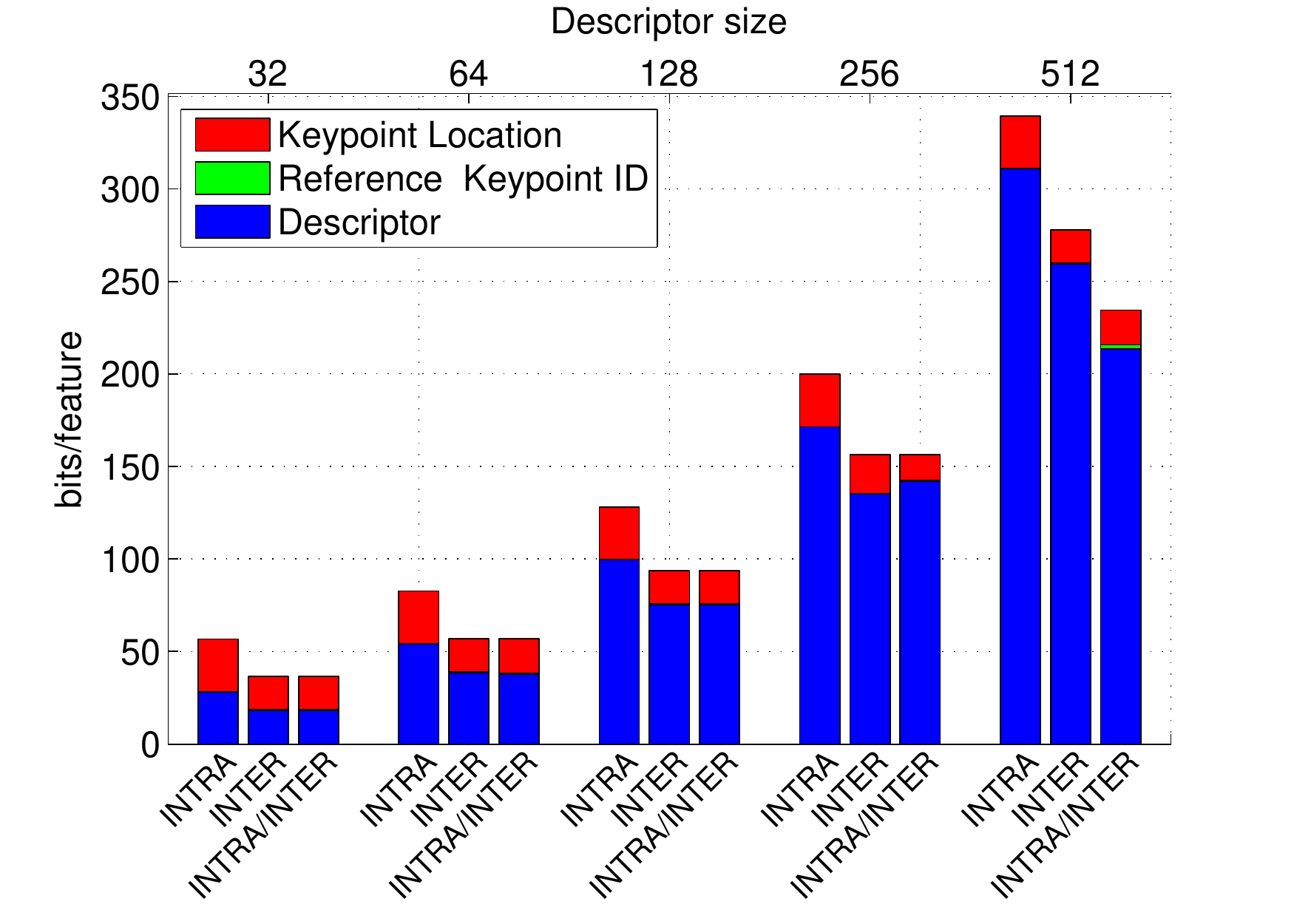}\label{fig:RD_ada_foreman}}
		\subfigure[]{\includegraphics[trim=0.7cm 0.3cm 1.5cm 0.1cm, clip=true,width=0.42\textwidth]{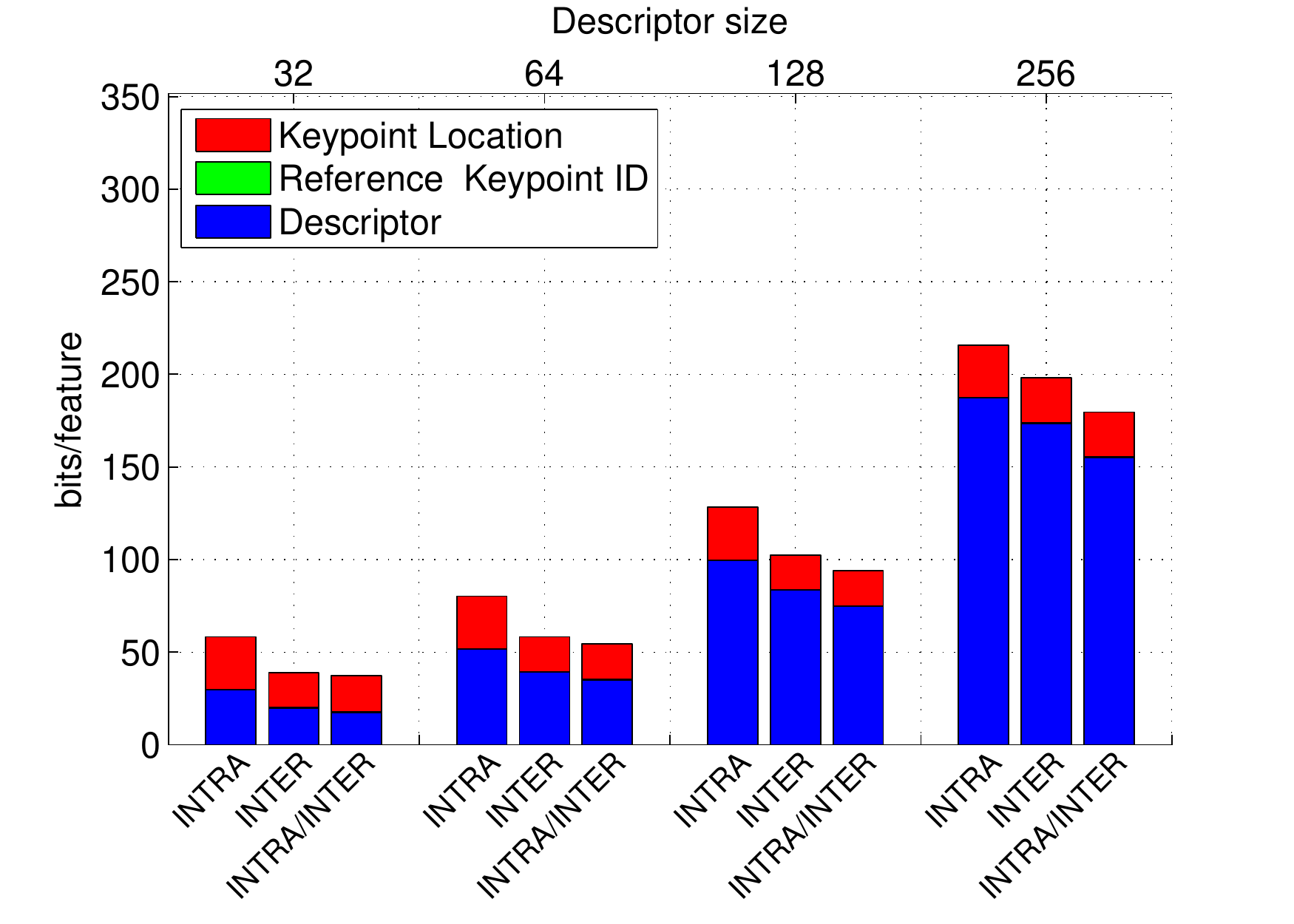}\label{fig:RD_foreman}}
		% \caption{Bitrate needed to encode each visual feature extracted from the \emph{Foreman} sequence, varying the size of the binary descriptor, for a) BRISK, with BAMBOO as dexel selection scheme; b) BINBOOST}
		\caption{Bitrate needed to encode each visual feature extracted from the \emph{Foreman} sequence, varying the size of the binary descriptor, for a) BRISK; b) BINBOOST.}
		\label{fig:rate_foreman} 
	\end{figure*}

	\begin{figure*}[t]
		\centering
		\subfigure[]{\includegraphics[trim=0.8cm 0.1cm 1.2cm 0.4cm, clip=true,width=0.42\textwidth]{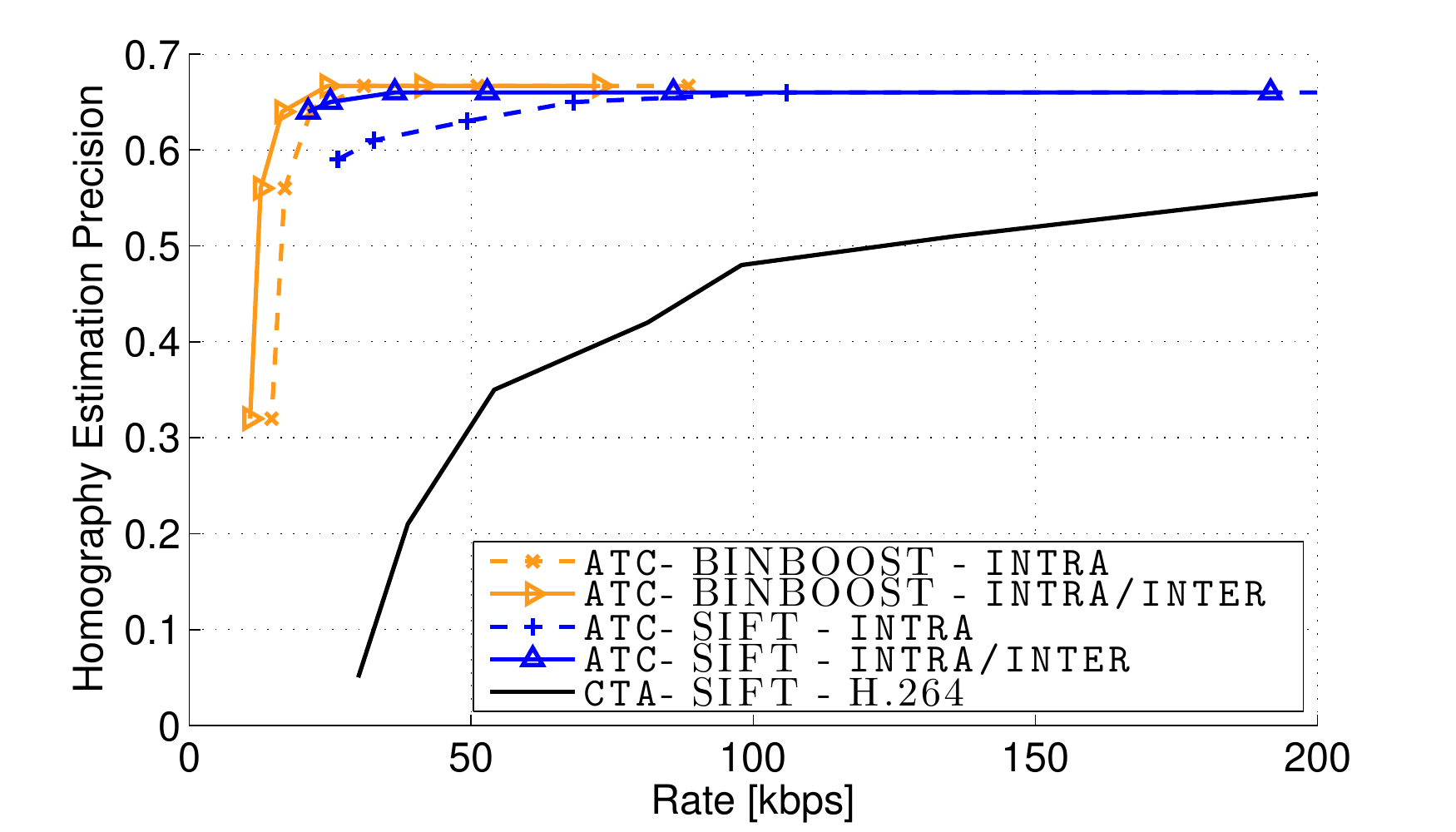}\label{fig:Paris_homog_BINBOOST}}
		\subfigure[]{\includegraphics[trim=0.8cm 0.1cm 1.2cm 0.4cm, clip=true,width=0.42\textwidth]{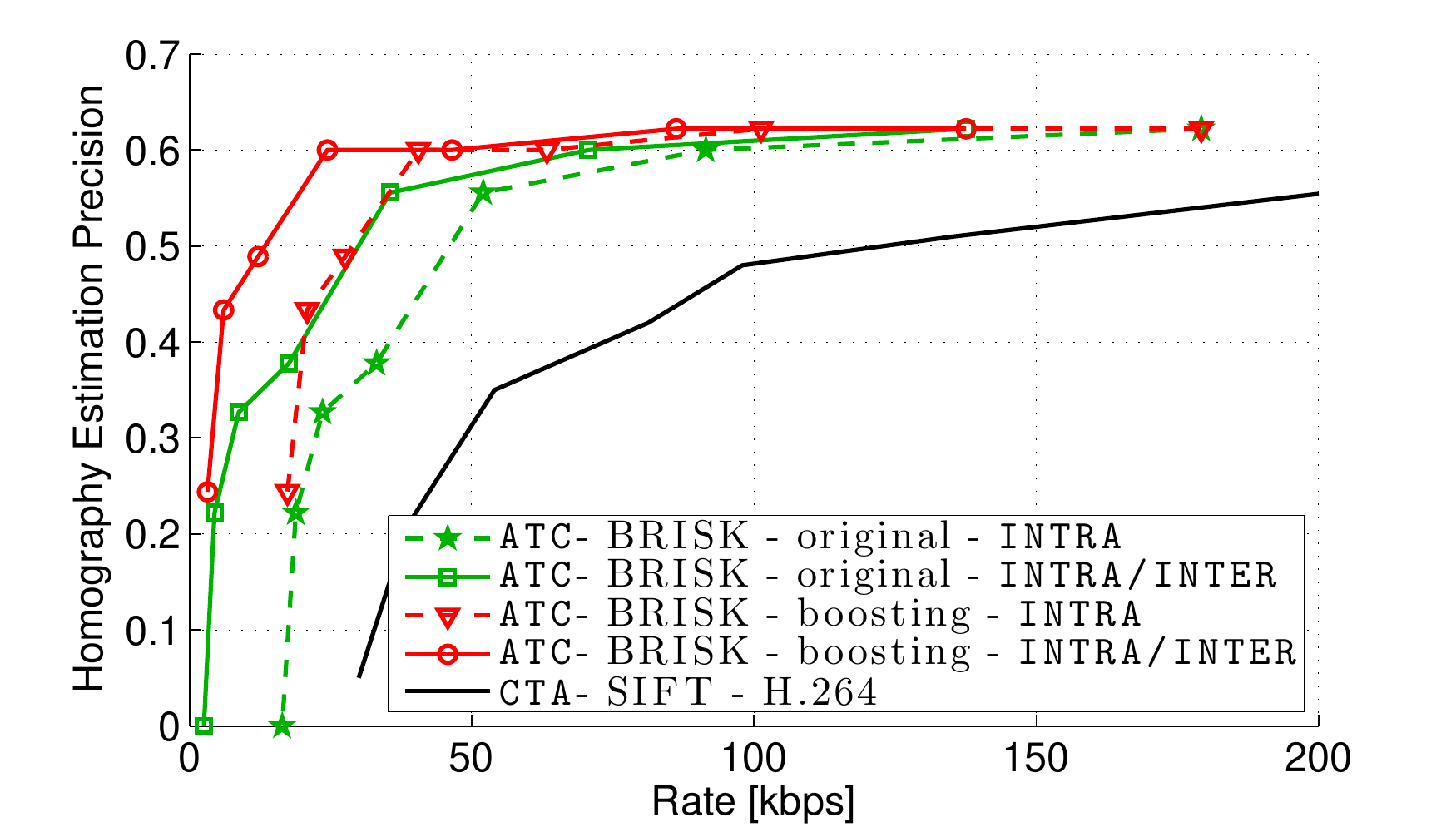}\label{fig:Paris_homog_BAMBOO}}
		\caption{Rate-accuracy curves obtained for the \emph{Paris - homography} sequence. a) ATC (either based on SIFT or BINBOOST) vs. CTA; b) BAMBOO boosted dexel selection scheme vs. BRISK original dexel selection scheme within the ATC approach.}
		\label{fig:homog_Paris} 
	\end{figure*}

	\subsubsection{Homography estimation}
	% local visual features traditionally represent an effective solution to such a task. 
	First, we evaluated the number of bits necessary to encode each visual feature using either intra-frame or inter-frame coding, when varying the size of the descriptor $K$. Figure~\ref{fig:rate_foreman} shows the bitrate obtained by coding the BRISK features extracted from \emph{Foreman} video sequence, indicating separately the number of bits used for encoding the keypoint location, the reference keypoint identifier (inter-frame only), and the descriptor elements. 
	% As for BRISK, we tested the dexel selection scheme based on asymmetric pairwise boosting presented in Section~\ref{sec:BAMBOO}.
	At high bitrates ($K = 256$), the coding rate is equal to 200 bits/feature and 222 bits/feature in the case of intra-frame coding, 156 bits/feature and 178 bits/feature in the case of intra-frame coding for BRISK and BINBOOST, respectively. At low bitrates ($K = 32$), the rate drops to approximately 55 bits/feature and 40 bits/feature for intra- and inter-frame coding, respectively. Similar results were also obtained for the other test sequences.

	Figure~\ref{fig:homog_Paris} compares the results of $\ATC$ and $\CTA$. As a benchmark, we also included the results obtained using $\ATC$ when SIFT visual features were used~\cite{BaroffioRCTT:TIP}. As a reference, when no visual feature compression is used, the bitrate for sending SIFT, BINBOOST or BRISK descriptors in the ATC paradigm would be, respectively, 376 kbps, 107 kpps and 220 kbps, attaining a homography estimation precision equal to 0.66, 0.66 and 0.62. Thus, visual feature compression leads to very large coding gains, since comparable precision levels are achievable with at approximately 25 kbps for SIFT, BINBOOST and BRISK (bitrate saving -93\%, -77\% and -89\%, respectively).
	In all cases, ATC outperformed CTA, since higher levels of precision are attained for all target bitrates. With respect to the $\ATC$ approach, inter-frame coding significantly improves the coding efficiency, especially at low bitrates.
	 % The use of SIFT in ATC allows to achieve a higher accuracy in the homography estimation, but at the cost of a significantly higher complexity to extract the visual features at the sensing node. This is particularly important in visual sensor network applications, in which sensing nodes are critically energy-constrained.

	In addition, to evaluate the benefit of using the dexel selection scheme described in Section~\ref{sec:BAMBOO}, we compared our results with a baseline in which the original selection scheme embedded in the BRISK descriptor was used. The latter simply chooses the elements corresponding to smallest spatial distance between the pattern points whose intensities are to be compared. Figure~\ref{fig:Paris_homog_BAMBOO} shows that appropriately selecting the dexels significantly improves the task accuracy, which saturates using as few as 64 dexels / descriptors (requiring approximately 25 kbps to be transmitted). 

	\begin{figure*}[t]
		\centering
		\subfigure[]{\includegraphics[trim=0cm 0cm 0cm 0cm, clip=true,width=0.47\textwidth]{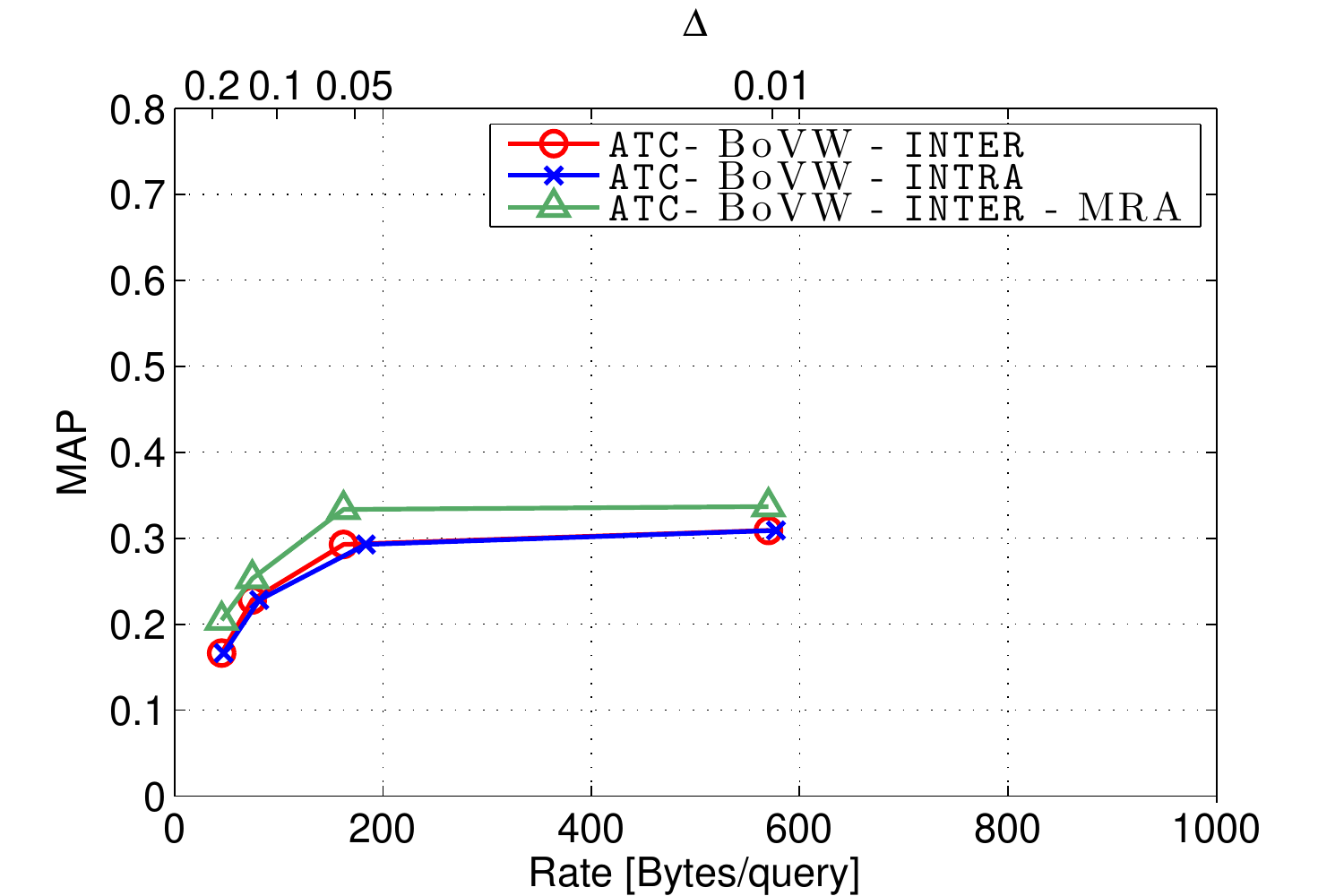}\label{fig:BOW4k}}
		\subfigure[]{\includegraphics[trim=0cm 0cm 0cm 0cm, clip=true,width=0.47\textwidth]{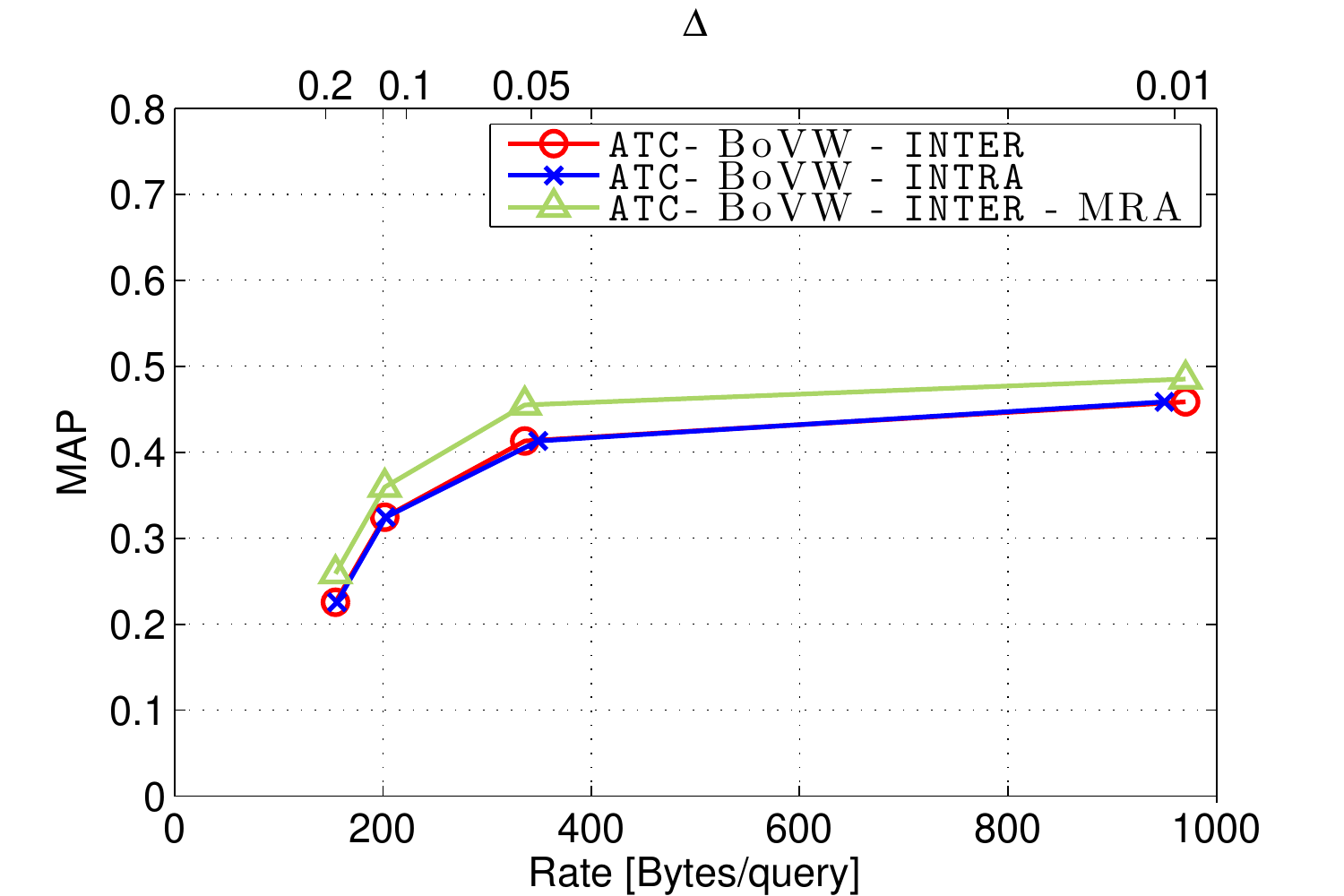}\label{fig:BOW16k}}
	 	\caption{Rate-MAP curves for the retrieval task, when matching is performed resorting to Bag-of-Words based on BRISK local features, considering a dictionary of a) $M = 4096$ visual words; b) $M = 16384$ visual words. 
	% ``Median Rank Aggregation (MRA)'' boosts the accuracy performance of the retrieval task. 
	}
		\label{fig:BOW} 
	\end{figure*}
	%MT si riesce a mettere la stessa scala sull'asse x?
	%MT Userei lo stesso tono di verde poi usato in Fig. 7
	%LB ok

	\subsubsection{Content-based retrieval task}

	Given a query video sequence, the task consists in retrieving the relevant images within a database composed of $Z = 10000$ images using global features and, possibly, refine the result using local features. 

	Since global features are computed from local features, we evaluated first the impact of the BRISK detection threshold, which determines the number of local features extracted from each query frame. A high threshold value leads to a low number of local features and, consequently, to sparser BoVW global descriptors. This allows for more efficient encoding, at the cost of less discriminating, and thus less accurate, global descriptors. In contrast, a sufficiently low threshold (high number of local features) allows unstable descriptors to be detected and leads to noisy global descriptors. Table~\ref{tab:MAP_M_th} shows the impact of both dictionary size and BRISK detection threshold on the \emph{Mean Average Precision} measure. A BRISK threshold value set to a value of 50 leads to the best results for all the possible dictionary sizes. 
	% Figure~\ref{fig:BOW_3D} shows the impact of both dictionary size and BRISK detection threshold on the \emph{Mean Average Precision} measure. A BRISK threshold value set to a value of 50 leads to the best results for all the possible dictionary sizes. 

	\begin{table}
		\begin{center}
			\caption{Mean Average Precision (MAP) for the retrieval task, as a function of the size of the number of visual words $M$ composing the dictionary and the BRISK detection threshold.}
	\begin{tabular}{cc|c|c|c|c|l} 
	\cline{3-6}
	& & \multicolumn{4}{ c| }{BRISK threshold} \\ \cline{3-6}
	& & 30 & 50 & 70 & 90 \\ \cline{1-6} 
	\multicolumn{1}{ |c }{\multirow{3}{*}{\# words} } &
	\multicolumn{1}{ |c| }{1k} & .15 & \textbf{.21} & .18 & .15 &     \\ \cline{2-6}
	\multicolumn{1}{ |c  }{}                        &
	\multicolumn{1}{ |c| }{4k} & .23 & \textbf{.31} & .28 & .21 &     \\ \cline{2-6}
	\multicolumn{1}{ |c  }{} &
	\multicolumn{1}{ |c| }{16k} & .30 & \textbf{.46} & .44 & .35 & \\ \cline{1-6}
	\end{tabular}
	\label{tab:MAP_M_th}
	\end{center}
	\end{table}

	% \begin{figure}[t]
	% 	\centering
	% 	\includegraphics[trim=0cm 0cm 0cm 0cm, clip=true,width=0.47\textwidth]{figures/MAP3D.pdf}\label{fig:homog}
	% 	\caption{Mean Average Precision (MAP) for the retrieval task, as a function of the size of the visual word dictionary $M$ and the BRISK detection threshold.}
	% 	\label{fig:BOW_3D} 
	% \end{figure}

	%MT forse è più chiaro mostrare questo con una tabellina, mettendo in bold i valori per la soglia pari a 50
	%LB ok

\begin{figure}[t]
		\centering
		\includegraphics[trim=0cm 0cm 0cm 0cm, clip=true,width=0.42\textwidth]{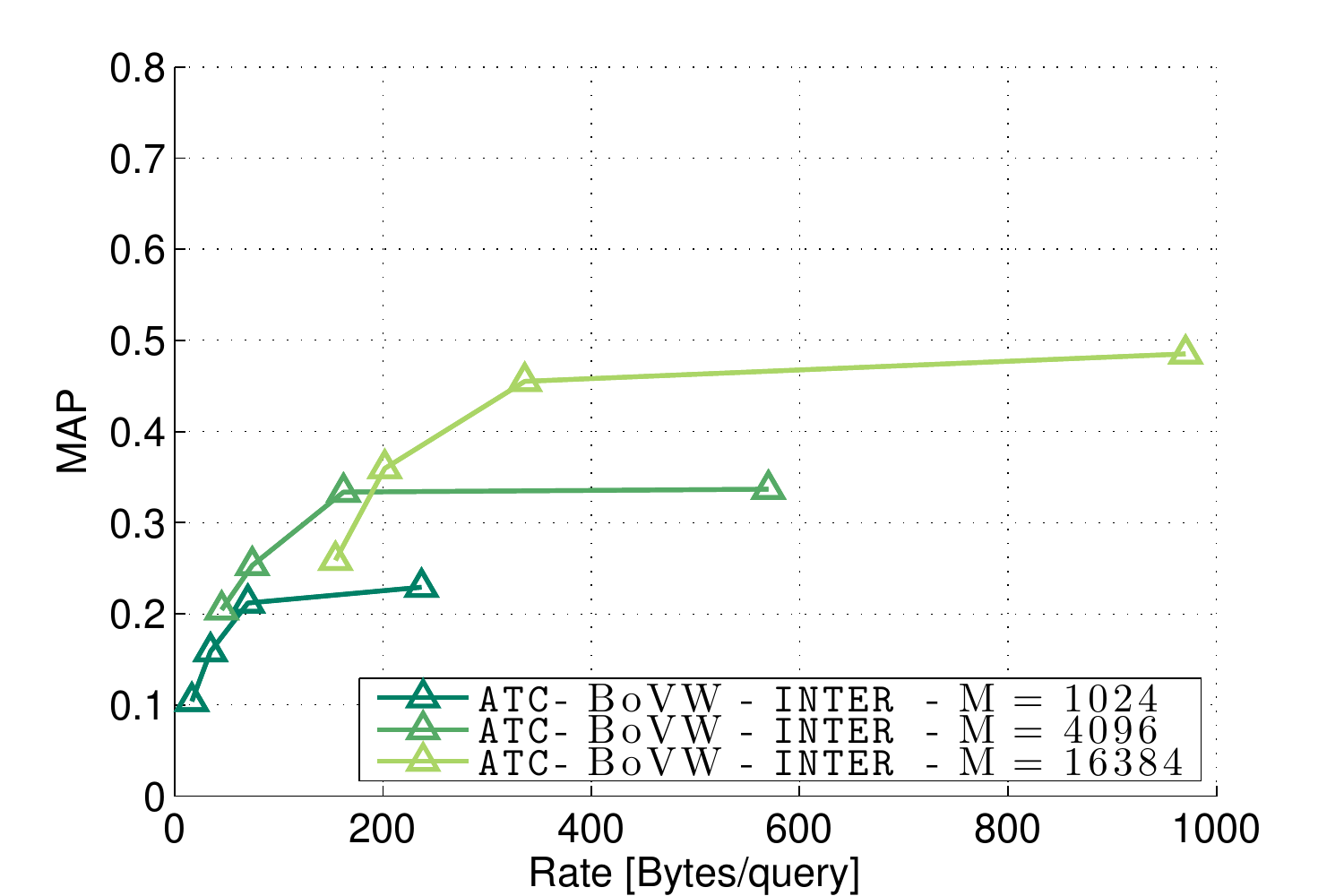}\label{fig:homog}
		\caption{Envelope of rate-MAP curves for the content-based retrieval task, when matching is performed resorting to Bag-of-Words based on BRISK local features. The curves are obtained by varying both the dictionary size $M$ and the quantization step size $\Delta$, when ``Median Rank Aggregation (MRA)'' is employed.}
		\label{fig:BOW_envelope} 
	\end{figure}

	Then, we considered the impact of coding global features in $\ATC$, by tracing the rate-MAP curves obtained for different dictionary sizes. For example, Figure~\ref{fig:BOW4k} and ~\ref{fig:BOW16k} show the rate-MAP curves obtained with dictionary of size $M = 4096$ and $M = 16384$, respectively. Each curve was obtained by varying the quantization step size $\Delta$. A larger dictionary allows for improved accuracy. In particular, MAP saturates at approximately 0.34 and 0.49 when the dictionary has size $M = 4096$ and $M = 16384$, respectively. On the other hand, a larger dictionary leads to larger descriptors and, thus, a higher number of bits is required for each query. In details, the value of MAP saturates when using approximately 160 (180) and 350 (360) Bytes/query for $M = 4096$ and $M = 16384$, respectively, when inter-frame (intra-frame) coding is the selected method. Large dictionaries lead to quantizing similar features of consecutive frames to different visual words, thus reducing the amount of temporal redundancy and preventing inter-frame coding to achieve significant coding gains. Regardless of the dictionary size, the usage of \emph{Median Rank Aggregation} leads to an improvement of about 5\% in terms of MAP.
	 % Nonetheless, temporally coherent frames allows for the usage of \emph{Median Rank Aggregation}, leading to a retrieval performance improvement of about 5\%.
	%
	Figure~\ref{fig:BOW_envelope} summarizes the best rate-MAP curve for each dictionary size in the same chart, including also the case $M = 1024$. By inspecting the envelope of the rate-MAP curves, it is possible to observe that the dictionary size should be adjusted based on the target bitrate, namely, $M =1024$ when using less than 50 Bytes/query, $M = 16384$ when using more than 200 Bytes/query, and $M = 4096$ in all other cases.
	%
	% the results, showing the envelope of several rate-MAP curves corresponding to different dictionary sizes. A dictionary with a thousand word is the only viable option when the bitrate is constrained under abount 50 Bytes/query, whereas $M = 16384$ represents the best solution when using more than 200 Bytes/query. When the target bitrate is constrained to be within 50 and 200 Bytes/query, a dictionary composed of 4k words is the best performing option. 

\begin{figure}[t]
		\centering
		\includegraphics[trim=0cm 0cm 0cm 0cm, clip=true,width=0.47\textwidth]{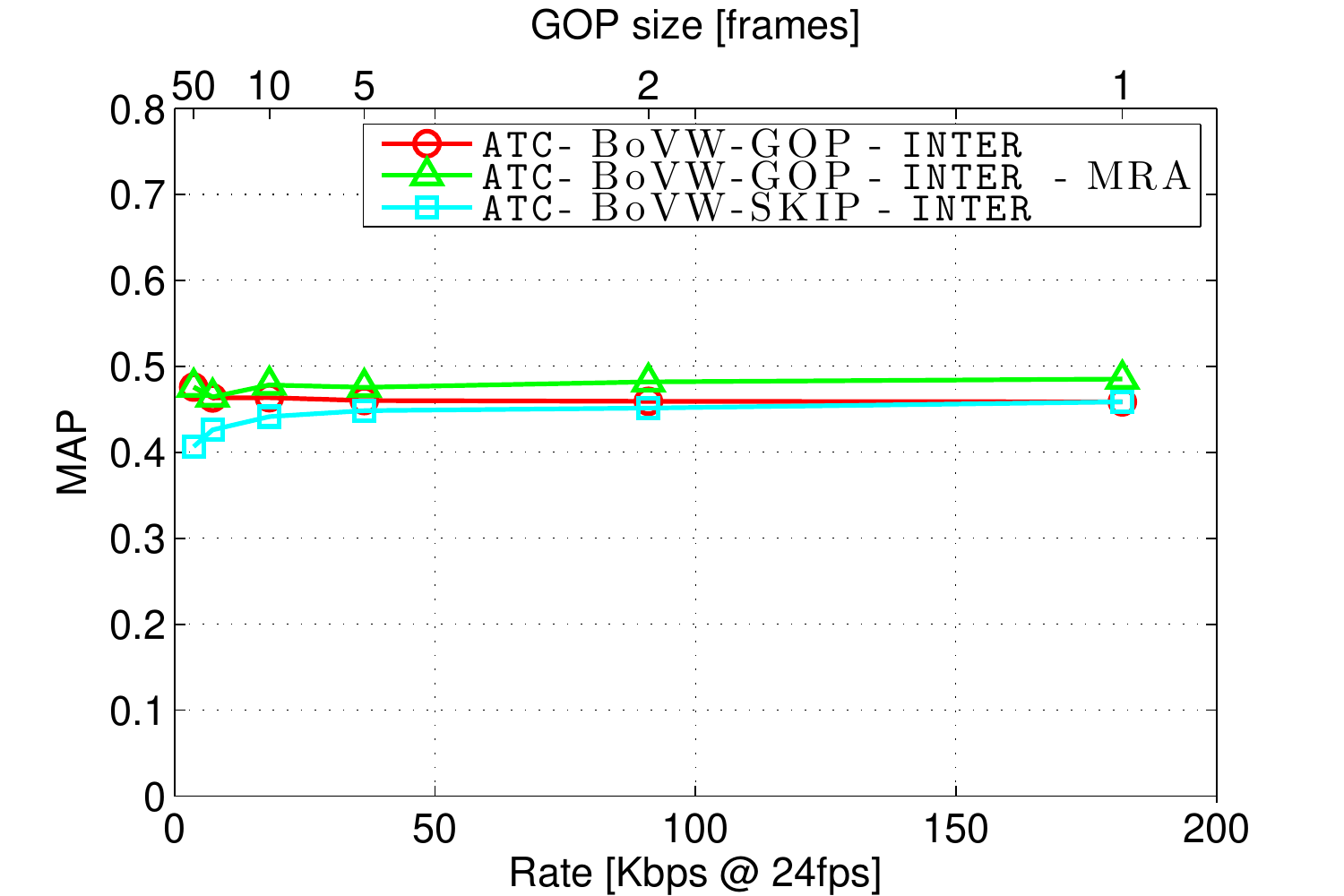}\label{fig:homog}
		\caption{Rate-MAP curves for the content-based retrieval task, when matching is performed using Bag-of-Words based on BRISK local features, considering a dictionary of $M = 16386$ visual words. } 
		% Global descriptors averaged within a group of pictures (GOP) provide better performance than global descriptors extracted from temporally sampled frames (SKIP), especially when considering large GOPs.}
		\label{fig:BOW_GOP} 
	\end{figure}

	As a further experiment, we fixed the dictionary size to $M = 16384$ to achieve the highest MAP, and we investigated how to reduce the rate by sending only one global descriptor per GOP, when the GOP size was varied in the set $\{1,2,5,10,20,50\}$. In Figure~\ref{fig:BOW_GOP} we observe that when using the BoVW-SKIP approach, the MAP slightly decreases when increasing the GOP size, while achieving a significant bitrate saving. This is due to the fact that fewer query frames were used for the same video query, thus reducing the bitrate but also the diversity in the query content. To overcome this issue, BoVW-GOP aggregates the global descriptors extracted from all frames of a GOP into a single descriptor. This leads to a significantly higher MAP (+8\%), while achieving the same bitrate saving. In addition, \emph{Median Rank Aggregation} can also be used at the receiver side to further improve the MAP. This is useful especially when considering small GOP sizes, i.e., when aggregation is performed resorting to a higher number of frames with a high temporal correlation. Although Figure~\ref{fig:BOW_GOP} might suggest that additional coding gains can be achieved by increasing the GOP size beyond 25 frames, in real application scenarios there are other requirements that typically constrain the largest GOP size allowed, namely the maximum tolerable delay, or the dynamic nature of the underlying video sequence. 

\begin{figure}[t]
		\centering
		\includegraphics[trim=0cm 0cm 0cm 0cm, clip=true,width=0.47\textwidth]{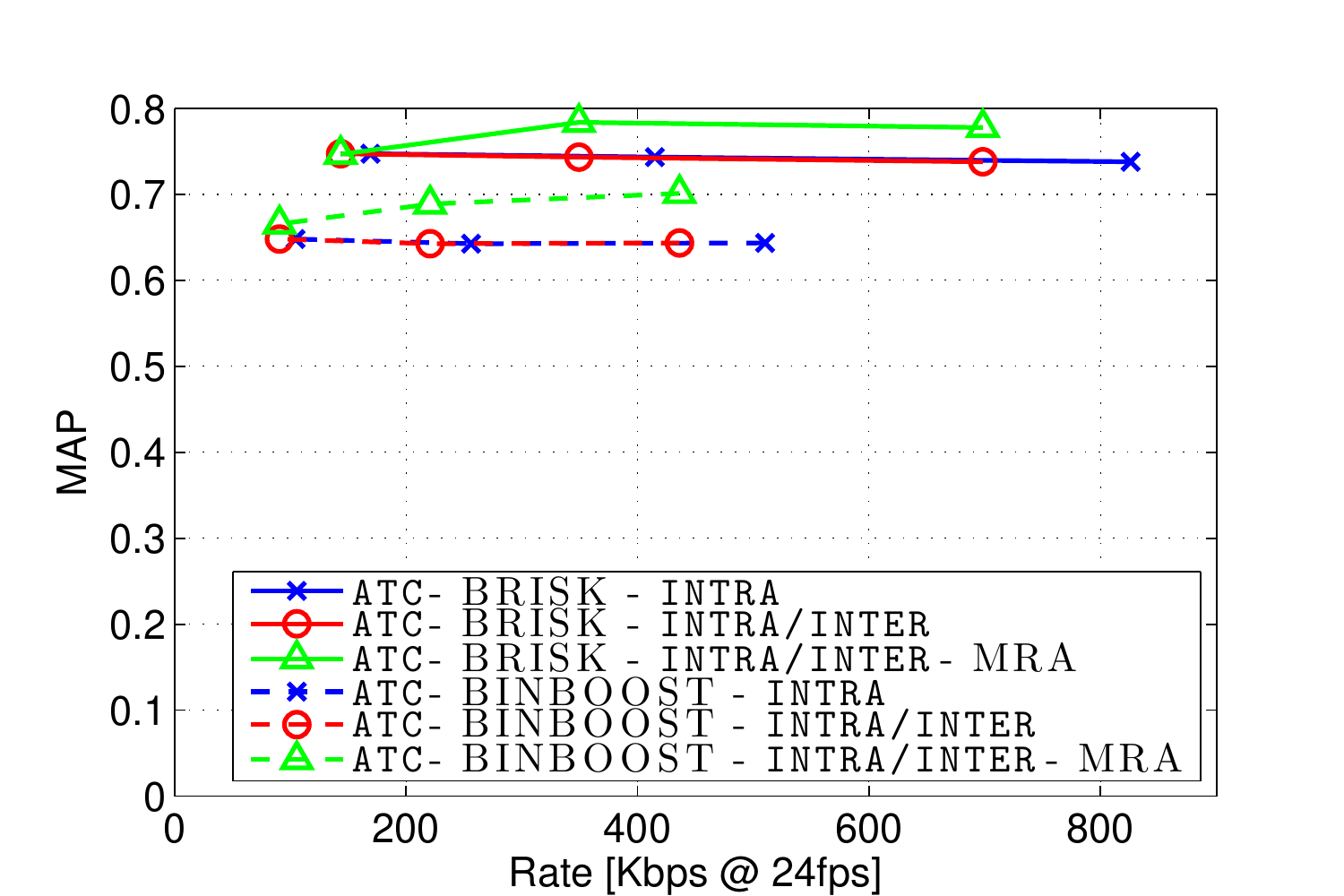}\label{fig:homog}
		\caption{Rate-MAP curves for the content-based retrieval task, when a re-ranking step is performed on the top-200 candidates, resorting to either BRISK or BINBOOST, as a function of the GOP size.}
		\label{fig:BRISK_retrieval} 
	\end{figure}

	In a typical content-based retrieval pipeline, local features are often used to re-rank the result obtained using global features. Figure~\ref{fig:BRISK_retrieval} shows the rate-MAP curves when either BRISK or BINBOOST descriptors were used in the re-ranking step. 
	% \hl{
	Similarly to the case of global descriptors, we investigated the impact of temporal subsampling on the overall accuracy. Considering a Group Of Pictures (GOP), a set of visual features is extracted from the first frame of such GOP and used in order to refine the results provided by the retrieval pipeline based on global descriptors. Each curve is traced by varying the GOP size in the set $\{5,10,25\}$ and using the largest descriptor size ($K = 512$ for BRISK and $K = 256$ for BINBOOST).
	% }
	%MT check
	%LB ok
	With respect to the retrieval based on global features only, MAP was boosted from 0.49 to 0.78 (BRISK) and 0.69 (BINBOOST). Note that, unlike for the homography estimation task, BRISK outperforms BINBOOST for this task. At the same time, this comes at an additional cost in terms of bitrate, which is increased by approximately an order of magnitude. For example, when the GOP size is equal to 25, the bitrate increases from 8 kbps (global features) to 150 kbps for BRISK and 95 kbps for BINBOOST. Figure~\ref{fig:BRISK_retrieval} also shows that inter-frame coding reduces the bitrate with respect to intra-frame coding between 5\% and 15\%, depending on the GOP size. Similarly to the case of global features, Median Rank Aggregation brings significant advantages in terms of MAP, when a sufficiently small GOP size is employed. 

\begin{figure}[t]
		\centering
		\includegraphics[trim=0cm 0cm 0cm 0cm, clip=true,width=0.47\textwidth]{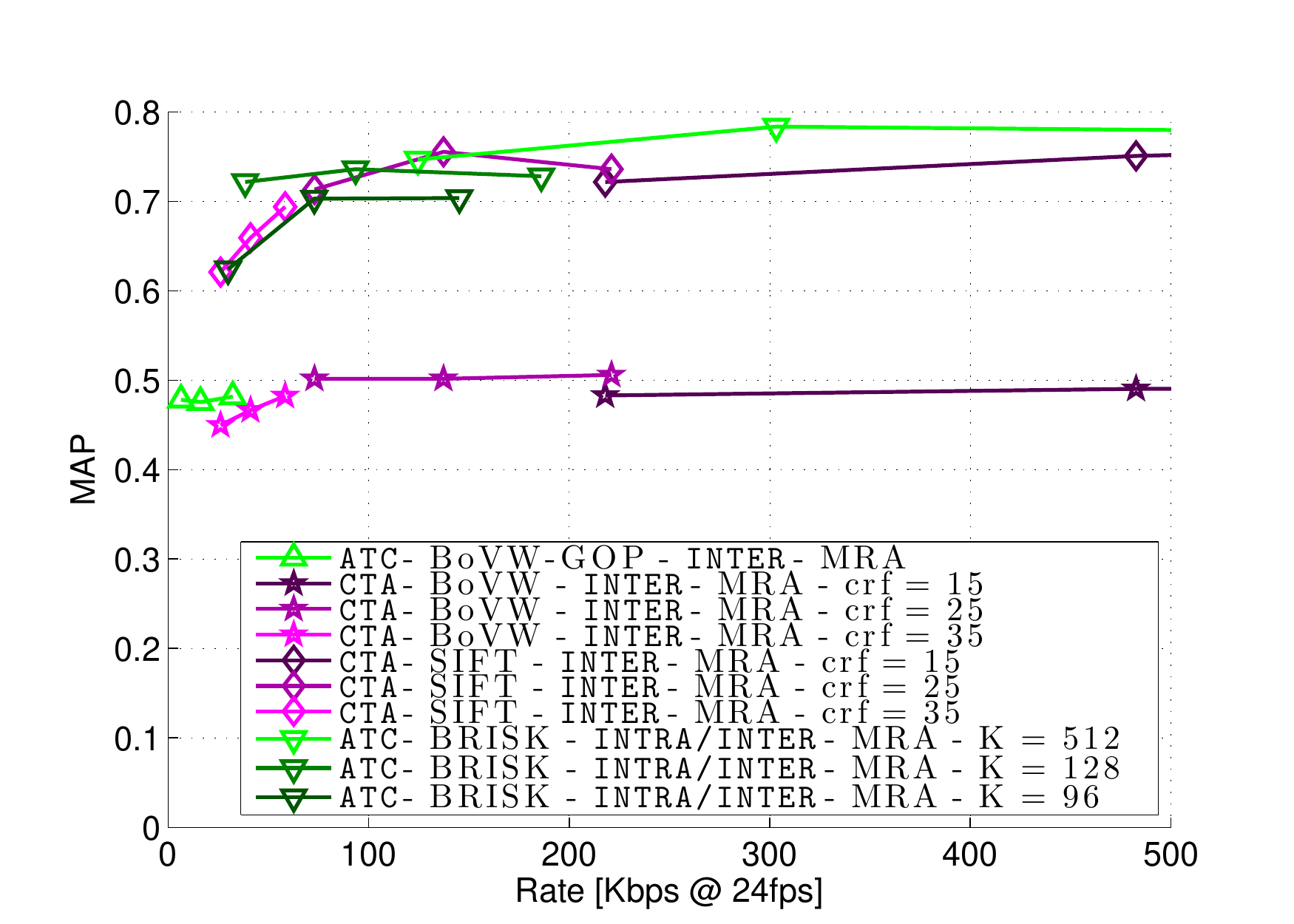}\label{fig:homog}
		\caption{Rate-accuracy curves comparing $\ATC$ and $\CTA$ approaches.}
		\label{fig:ATC_CTA_full} 
	\end{figure}

	Finally, we compared the results obtained resorting to either $\ATC$ or $\CTA$ in Figure~\ref{fig:ATC_CTA_full} (note that the curve $\ATC-BoVW$ corresponds to the operating points in the MAP-rate curve in Figure~\ref{fig:BOW_GOP} corresponding to a GOP size equal to either 25, 10 or 5 frames). When using global features only, $\ATC$ outperforms $\CTA$ by a large margin. Indeed, at very low bitrate, $\ATC$ based on global features is the only viable option, since at least 30kbps are needed to transmit a pixel-level representation of the visual content and thus, to enact the $\CTA$ paradigm. 
	%MT ho un dubbio. Se il GOP size del punto più a sx è veramente 50, vuol dire che abbiamo un frame ogni due secondi. Questo frame ci costa veramente 60.000 bit? 
	%LB si, con descrittori a 512 bit. 256 feature * 512 bit * 0.5 (coding ratio) ~ 60kb. Utilizzando descrittori a 128 bit tale costo si riduce a ~17kb, riducendo la MAP di 3 punti. 

	When setting the GOP size to 10 frames (i.e., corresponding to the operating point in the middle of each curve), $\ATC$ requires as few as 18 kbps to achieve a MAP equal to 0.48. In contrast, $\CTA$ requires 40 kbps (MAP = 0.46), 140 kbps (MAP = 0.50) and 480 kbps (MAP = 0.49), when changing the Constant Rate Factor parameter $\textit{crf}$ of H.264/AVC. 

	When considering re-ranking based on local features, $\CTA$ is able to significantly improve MAP at no extra cost in terms of bitrate. The best performance achieved by $\CTA$ at $\textit{crf} = 25$ (for both global and local features) can be attributed to the mild smoothing operated by lossy coding at this bitrate, which reduces noise and allows detecting more stable keypoints. Conversely, $\ATC$ requires sending additional bits to be able to encode the local features. Figure~\ref{fig:ATC_CTA_full} shows different curves obtained by varying the number of dexels $K$. 
	% \hl{
	In particular, descriptors with size equal to 512, 128 or 96 dexels were tested. Smaller descriptor lengths lead to a significant loss in terms of accuracy. This is due to the inefficiency of a very short BRISK descriptor.
	% , which might be overcome by the design of future binary descriptors.
	% } 
	In the case of local descriptors, $\ATC$ performs on a par with $\CTA$, and what is the best paradigm is determined by the target bitrate. For example, at 40 kbps, MAP is equal to 0.72 for $\ATC$ and 0.65 for $\CTA$. Conversely, at 30 kbps, MAP is equal to approximately 0.63 for both $\ATC$ and $\CTA$. 
	%MT hai mai provato con una dimensione tra 64 e 128? Non sia mai che si riesca a stare sopra la curva CTA crf = 25
	%LB TODO

	%MT toglierei la curva K = 256. Aggiunge un po' di confusione, ma non mi sembra che aggiunga punti di lavoro interessanti.
	%LB ok

\section{Conclusions}\label{sec:conclusions}

We proposed two coding architectures tailored to either local binary features (tested on BRISK and BINBOOST) or global features (based on Bag-of-Visual-Words), extracted from video sequences. The efficiency of the proposed solution was evaluated by means of rate-efficiency curves with respect to traditional visual analysis tasks. In the case of homography estimation the $\ATC$ paradigm always outperforms $\CTA$ by a large margin, achieving the same task efficiency that can be obtained using uncompressed sequences with as few as 20 kbps. In the case of content-based retrieval, the $\ATC$ paradigm always outperforms $\CTA$ when using global features, operating at 8 kbps and achieving the same MAP obtained using uncompressed sequences. When using local features, $\ATC$ and $\CTA$ perform on a par, calling for the investigation of more compact descriptors and more sophisticated coding tools (e.g., filtering the keypoints to be encoded based on the temporal coherence). 
Future work will address the use of recently proposed global descriptors extracted from binary features, e.g. BVLAD~\cite{ICIP2014:Steinbach:BVLAD}, and hybrid $\CTA-\ATC$ coding schemes.

\IEEEpeerreviewmaketitle

\ifCLASSOPTIONcaptionsoff
  \newpage
\fi

\bibliographystyle{IEEEtran}
\bibliography{main}

\end{document}